\documentclass[pre,aps,twocolumn,amsmath,amssymb]{revtex4-1}

\pdfoutput=1
\usepackage{hyperref}
\usepackage{graphicx}
\usepackage[usenames]{color}
\usepackage{bm}
\usepackage[final]{movie15}

\def\cal#1{\mathcal{#1}}
\def\eqq#1{Eq.~(\ref{#1})}
\def\eq#1{(\ref{#1})}
\def\av#1{\langle #1 \rangle}

\def\f#1{Fig.~\ref{#1}}

\def\c#1{~\cite{#1}}

\def\x{{\bm x}}

\def\s#1{Section~\ref{#1}}

\def\e{{\rm e}}

\def\beq{\begin{equation}}
\def\eeq{\end{equation}}
\def\bea{\begin{eqnarray}}
\def\eea{\end{eqnarray}}

\begin{document}

\title{Sampling rare fluctuations of discrete-time Markov chains}
\author{Stephen Whitelam}\email{{\tt swhitelam@lbl.gov}}
\affiliation{Molecular Foundry, Lawrence Berkeley National Laboratory, 1 Cyclotron Road, Berkeley, CA 94720, USA}
\begin{abstract}
We describe a simple method that can be used to sample the rare fluctuations of discrete-time Markov chains. We focus on the case of Markov chains with well-defined steady-state measures, and derive expressions for the large-deviation rate functions (and upper bounds on such functions) for dynamical quantities extensive in the length of the Markov chain. We illustrate the method using a series of simple examples, and use it to study the fluctuations of a lattice-based model of active matter that can undergo motility-induced phase separation.
\end{abstract}

\maketitle

\section{Introduction}
\label{intro}

Quantifying the rare behavior of stochastic models is necessary to understand important physical processes such as phase separation and chemical reactions\c{frenkel2001understanding}. Computer simulation methods for quantifying rare behavior include umbrella sampling\c{torrie1977nonphysical}; forward-flux sampling\c{allen2009forward}; transition-path sampling\c{bolhuis2002transition}; and diffusion Monte Carlo methods\c{giardina2011simulating,giardina2006direct,nemoto2016population,lecomte2007numerical,nemoto2016iterative}, which often involve the use of an auxiliary or reference model whose typical behavior is in some sense equivalent to the rare behavior of the model of interest\c{bucklew1990large,touchette2009large,garrahan2009first,maes2008canonical,giardina2011simulating,garrahan2009first,lecomte2007numerical,jack2015effective,nemoto2014computation,nemoto2016population}. Motivated by this latter approach, we show that the rare fluctuations of a discrete-time Markov chain generated by a stochastic model of interest can be quantified in a simple way using a stochastic {\em reference model} whose parameters are simply related to those of the original model. Partial information about the original model's rare fluctuations -- meaning a bound on the large-deviation rate function for time-extensive observables -- can be obtained from typical trajectories of the reference model, while complete information about the original model's rare fluctuations can be obtained by considering constrained fluctuations of the reference-model trajectory ensemble. In Ref.\c{klymko2017rare} we use this procedure to study the rare behavior of models of growth; here we focus on Markov chains with well-defined steady-state measures. 

In what follows we describe the method, and use a series of simple examples to illustrate the method's application to Markov chains with well-defined steady-state measures (\s{sec_fluc}). We then apply the method to the lattice-based model of active matter introduced in Ref.\c{lattice1}, in order to quantify that model's rare dynamic fluctuations (\s{sec_active}). We conclude in \s{sec_conc}.

\section{Sampling the rare fluctuations of discrete-time Markov chains}
\label{sec_fluc}

Consider a discrete-time Markov chain of $K$ steps. At a given instant the system is in microstate (or `state') $C$, and with probability $p(C \to C')$ moves to state $C'$ (which may be $C$). Probability conservation requires that $\sum_{C'} p(C \to C')=1$. A trajectory $\x$ is a sequence of microstates $\x = \{C_0,C_1,C_2,\dots,C_K\}$ generated by this dynamics, and its probability is 
\beq
P[\x]  = P_0(C_0) \prod_{k=0}^{K-1} p(C_k \to C_{k+1}),
\eeq 
where $P_0(C_0)$ is the probability of starting in state $C_0$. The master equation for this discrete Markov chain is
\beq
\label{me1}
{\bm P}(k+1) = {\bm W} {\bm P}(k),
\eeq
where ${\bm P}(k)$ is the column vector whose elements are $P(i,k)$, the probability of residing in state $i$ after $k$ steps of the dynamics, and ${\bm W}$ is the matrix whose $(j,i)^{\rm th}$ element is $p(i \to j)$. For an $M$-state system, ${\bm P}$ has $M$ elements and ${\bm W}$ has $M \times M$ elements.

Consider now an observable $A[\x]$, called the `activity', extensive in the length of the trajectory\c{garrahan2009first}. Upon moving from $C$ to $C'$ we increment $A$ by an amount $\alpha(C\to C')$, so that $A[\x] = \sum_{k=0}^{K-1} \alpha(C_k \to C_{k+1})$. Our aim is to compute the probability
 \beq
 \label{sum}
 \rho(a,K) \equiv \sum_{\x} P[\x] \delta{(A[\x]-K a)}
 \eeq 
 that a trajectory of length $K$ possesses a particular value $a \equiv {A}/K$ of the intensive counterpart of the observable $A$. For many models this probability adopts a large-deviation form $\rho(a,K) \sim \e^{-K I(a)}$ for large values of $K$, where $I(a)$ is the large-deviation rate function\c{touchette2009large}. \eqq{sum} is the instruction to simulate the original model and count the trajectories that possess activity $A=Ka$. When $a$ is close to a `typical' value $a_0$, for which $I(a_0)=0$, such sampling is efficient. But when $a$ is far from $a_0$, so that it is rarely generated by direct simulation, a different strategy is required.

To motivate this strategy, note that the desired quantity $\rho(a,K)$ can be regarded as the normalization factor (or `partition function') of the {\em microcanonical path ensemble}\c{evans2004rules,evans2004detailed,chetrite2013nonequilibrium,Chetrite2014}
\beq
P_a[\x] = \frac{P[\x] \delta{(A[\x]-K a)}}{\rho(a,K)},
\eeq
which defines an ensemble of trajectories conditioned upon the constraint $A[\x] = Ka$. (The term `microcanonical' comes from analogy with the fixed-energy ensemble of equilibrium statistical mechanics; here it is activity that is fixed). Alternatively one can define a {\em canonical path ensemble} (or tilted ensemble or $s$-ensemble)\c{touchette2009large} in which $A[\x]$ can fluctuate,
\beq
\label{se}
P_s[\x] =\frac{\e^{-s A[\x]} P[\x] }{\sum_\x \e^{-s A[\x] }P[\x]}.
\eeq
The master equation in the $s$-ensemble is
\beq
{\bm P}_s(k+1) = {\bm W}_s {\bm P}_s(k),
\eeq
where ${\bm W}_s$ is the matrix whose $(j,i)^{\rm th}$ element is 
\beq
p_s(i \to j) =\e^{-s \alpha(i \to j)} p(i \to j).
\eeq
 From the canonical path ensemble one can in certain cases recover the desired quantity $\rho(a,K)$\c{chetrite2013nonequilibrium,Chetrite2014}. The logarithm of the normalization factor in \eqq{se}, $\lambda(s) \equiv K^{-1} \ln \sum_\x \e^{-s A[\x] }P[\x]$, can be obtained by taking the logarithm of the principal eigenvalue of the generator ${\bm W}_s$ of $s$-ensemble dynamics\c{touchette2009large,garrahan2009first}. Legendre transform of $\lambda(s)$ yields a rate function
\beq
\label{leg}
I_s(a) = \max_s(-sa-\lambda(s)),
\eeq
which is equal to the true rate function $I(a)$ when the latter is convex\c{dinwoodie1992large,dinwoodie1993identifying,touchette2009large}. The $s$-ensemble has been used with great success, providing insight into the behavior of a large number of models\c{garrahan2009first}, but it possesses two drawbacks. One is that ${\bm W}_s$ is not a stochastic generator (its columns do not sum to unity, i.e. $\sum_j \e^{-s \alpha(i \to j)} p(i \to j)\neq 1$), and so considerable effort and ingenuity is sometimes needed to identify and simulate the $s$-ensemble\c{garrahan2009first,touchette2009large,maes2008canonical,giardina2011simulating,lecomte2007numerical,jack2015effective,nemoto2014computation,nemoto2016population}. More seriously, the calculation of probabilities is indirect, and fails when the rate function $I(a)$ of the original model is non-convex\c{dinwoodie1992large,dinwoodie1993identifying,touchette2009large}. This is so in many physically interesting cases, where there exist phase transitions and coexistence of trajectories. Here \eq{leg} returns only the convex hull of the true rate function\c{touchette2009large}: information about the trajectory ensemble is missing.

The method described here and in Ref.\c{klymko2017rare} draws inspiration from the $s$-ensemble method but calculates $\rho(a,K)$ directly, and so can recover non-convex rate functions. It is related to the method used in Ref.\c{rohwer2015convergence}, generalized to Markov chains and applied to a probability-conserving model. Consider a stochastic reference model whose transition probabilities $p_{\rm ref}(i \to j)$ are normalized versions of those of the $s$-ensemble, $p_s(i \to j)/\sum_j p_s(i \to j)$ (the idea of using such a model is noted in Eq. (4.24) of\c{touchette2011basic}). The master equation of the reference model is then
\beq
\label{me2}
{\bm P}_{\rm ref}(k+1) = {\bm W}_{\rm ref} {\bm P}_{\rm ref}(k),
\eeq
where element $(j,i)$ of ${\bm W}_{\rm ref}$ is 
\beq
p_{\rm ref}(i \to j) = \frac{\e^{-s \alpha(i \to j)} p(i \to j)}{\sum_j \e^{-s \alpha(i \to j)} p(i \to j)}.
\eeq
Note that the columns of ${\bm W}_{\rm ref}$ sum to unity --  $\sum_j p_{\rm ref}(i \to j) =1$ -- and so ${\bm W}_{\rm ref}$ is stochastic (probability-conserving). The idea of the scheme is then as follows. The reference model is not the $s$-ensemble, and is not intended to be. Simulations of the reference model (parameterized by $s$) produce values of $a$ that for large $K$ concentrate on $a_s$, where $a_s$ minimizes the rate function $I_{\rm ref}(a)$ of the reference model. Carrying out such simulations, and using the equations shown below, we calculate a piece $I(a_s)$ of the rate function of the original model. Repeating the process for a different value of $s$ yields another piece of $I(a)$, and so on. We do not generate the microcanonical or canonical path ensembles. Instead, a simple modification of the original model (which admits all the processes of that model and no new ones) allows us to control which value of $a$ is sampled, and from comparison of original and reference models we calculate the likelihood, \eqq{sum}, of generating those values of $a$ using the original model.

To carry out this calculation note that \eqq{sum} can be written
\bea
\label{summ}
\rho(a,K) =\sum_\x P_{\rm ref}[\x] w[\x] \delta{(A[\x]-Ka)},
\eea
where 
\beq
P_{\rm ref}[\x]  = P_0(C_0) \prod_{k=0}^{K-1} p_{\rm ref}(C_k \to C_{k+1})
\eeq 
 is the trajectory weight of the reference model, and $w[\x] \equiv P[\x]/P_{\rm ref}[\x]$. It is convenient to write the transition probabilities of the original and reference models as $p(C \to C') = W(C \to C')/R(C)$ and $p_{\rm ref}(C \to C') = W_{\rm ref}(C \to C')/R_{\rm ref}(C)$, where $W(C \to C')$ and 
 \beq
\label{rm}
W_{\rm ref}(C \to C') = {\rm  e}^{-s \alpha(C \to C')} W(C \to C')
\eeq
are rates for the original and reference models, respectively, and $R(C) \equiv \sum_{C'} W(C \to C')$ and $R_{\rm ref}(C) \equiv \sum_{C'} W_{\rm ref}(C \to C')$ are the models' escape rates from a given state. 

We then have $w[\x] = \prod_{k=0}^{K-1} w_k$, where
\beq
\label{wi}
w_k= \e^{s \alpha(C_k \to C_{k+1})} \frac{R_{\rm ref}(C_k)}{R(C_k)}.
\eeq
It is also convenient to define the quantity
\beq
\label{qi}
q[\x] \equiv K^{-1} \sum_{k=0}^{K-1} \ln \frac{R_{\rm ref}(C_k)}{R(C_k)},
\eeq
using the pieces of $w[\x]$ not fixed by the delta-function path constraint. 

The sum in \eqq{summ} is the instruction to simulate the reference model, and take the arithmetic mean of values of $w$ for trajectories that display activity $a$ (the constraint imposed by the delta function). It is natural to consider the case $a=a_{\rm s}$, where $a_s$ is a value typical of the reference model (for which $I_{\rm ref}(a_s)=0$). If we generate ${\cal N}$ trajectories of the reference model, and the trajectories labeled $i=1,2,\dots, {\cal M} \leq {\cal N}$ are typical in this way, then \eq{summ} reads 
\bea
\label{inter}
\rho(a_s,K) = \frac{{\cal M}}{{\cal N}}\cdot \frac{1}{{\cal M}} (w_1 + \cdots + w_{\cal M}).
\eea
From \eq{wi} and \eq{qi} we have $w_\kappa= \e^{s K a_s} \e^{K q_\kappa}$, where $\kappa $ labels trajectories. Upon taking logarithms of \eq{inter} and assuming $K$ large, so that $K^{-1} \ln ({\cal M}/{\cal N})$ is negligible, we get
\beq
\label{ld2}
I(a_s) =- s a_s -q_s-K^{-1} \ln \int {\rm d}q\, P_s(q|a_s) \e^{K \delta q}.
\eeq
Here $I(a_s) = -K^{-1} \ln \rho(a_s,K)$ is the large-deviation rate function of the original model evaluated at $a_s$; $q_s \equiv \int {\rm d}q\, q \, P_s(q|a_s)$ is the mean value of $q[\x]$ for the ensemble of typical reference-model trajectories (those with $A[\x] = K a_s$); $P_s(q|a_s)$ is the probability distribution of $q$ for the ensemble of typical reference-model trajectories; and $\delta q \equiv q -q_s$. 

By Jensen's inequality the first two terms in \eq{ld2} provide a bound on the rate function $I_0(a_s) \geq I(a_s)$, where
\beq
\label{bound}
I_0(a_s) = -s a_s -q_s.
\eeq
 The full rate function can be recovered by evaluation of the integral, which is demanding in general\c{rohwer2015convergence} (e.g. requiring a high-order cumulant expansion) but straightforward when $P_s(q|a_s)$ is Gaussian or close to it (this is not guaranteed in general, but is realized in several cases we have encountered\c{klymko2017rare}).
 
 The computational cost ${\cal C}_{\rm ref}^{(0)}K $ of generating a trajectory of length $K$ of the reference model is in general greater than the cost ${\cal C}^{(0)}K$ of generating a trajectory of similar length using the original model, because to use the reference model we must calculate the bias parameters $\alpha(C \to C')$. However, this cost is usually more than offset by our ability to direct the reference model to rarely-sampled values $a_s$ of the dynamic parameter: the cost to sample the piece $I(a_s)$ of the rate function using the original model is $ {\cal C} ={\cal C}^{(0)}K \e^{K I(a_s)}$, while the cost using the reference model is ${\cal C}_{\rm ref}={\cal C}_{\rm ref}^{(0)}K \e^{K I_{\rm ref}(a_s)} = {\cal C}_{\rm ref}^{(0)}K$. Under most conditions we have encountered, the reference model is much more efficient than the original model. E.g. for the lattice model of \s{sec_active}, $\alpha(C \to C')$ can be calculated via spatially local updates (updates involving only the neighborhood of the chosen lattice site), and the ratio ${\cal C}_{\rm ref}^{(0)}/{\cal C}^{(0)}$ does not depend on the size of the system (it is of order 10). In this case the ratio ${\cal C}/{\cal C}_{\rm ref} =  ({\cal C}^{(0)}/{\cal C}_{\rm ref}^{(0)}) \e^{K I(a_s)}$ becomes large for any appreciable $K$ (and nonzero $I(a_s)$), and so the reference-model method is more efficient than direct simulation. If for instance the computation of the bias $\alpha(C \to C')$ required updates of the whole lattice, and so ${\cal C}_{\rm ref}^{(0)}$ grew with system size $N$, then there would exist an interval of $K$ for which direct simulation may be more efficient.

In what follows we specialize the discussion to ergodic (irreducible, aperiodic) Markov chains possessing well-defined steady-state measures (probability of occupancy) $\pi(C)$ and  $\pi_{\rm ref}(C)$. These are obtained by requiring \eq{me1} and \eq{me2} to be stationary, and satisfy
\beq
\pi_{(\rm ref)}(C) = \sum_{C'} \pi_{(\rm ref)}(C') p_{(\rm ref)}(C' \to C).
\eeq
The existence of such a measure is guaranteed when each microscopic process has a non-vanishing probability of reversal, which is true of the model of Ref.\c{lattice1}, but not true, for instance, of irreversible growth processes\c{klymko2017rare}. 

\eqq{ld2} can be written
\bea
\label{rf1}
I(a_s) = - K^{-1} \ln \av{\exp(\ln w_1 + \cdots + \ln w_K)}_{\rm ref},
\eea
where $\av{\cdot}_{\rm ref}$ denotes the average over the reference-model dynamics (for trajectories such that $a=a_s$). By Jensen's inequality we have \eqq{bound}, or
\bea
\label{bound2}
I_0(a_s) &=& -K^{-1} \ln \exp(\av{\ln w_1 + \cdots + \ln w_K}_{\rm ref}) \nonumber \\
&=& -\av{\ln w}_{\rm ref}.
\eea
The average $\av{\ln w}_{\rm ref}$ can be calculated by keeping track of the terms $\ln w_i$ associated with each move in a reference-model simulation,
\beq
\label{int0}
\theta(K) = K^{-1}\sum_{k=0}^{K-1} \ln w_k.
\eeq
For large $K$, the series $\theta(K)$ will converge to $\theta_s$, where
\bea
\label{int1}
\theta_s &=&  s a_s + q_s \\
\label{int2}
&=& a\sum_C \pi_{\rm ref}(C) \sum_{C'} p_{\rm ref}(C \to C') \alpha(C \to C') \nonumber \\&+&  \sum_C \pi_{\rm ref}(C) \ln \frac{R_{\rm ref}(C)}{R(C)}  \\
\label{int3}
&=&\sum_C \pi_{\rm ref}(C)  \left\{ -s\frac{\partial}{\partial s} \ln R_{\rm ref}(C)+\ln \frac{R_{\rm ref}(C)}{R(C)} \right\}.\hspace{0.4cm}
\eea
Note that for the choice $\alpha(C \to C') = B(C')$, made in Ref.\c{lattice1}, we have 
\bea
a_s &=& \sum_C \pi_{\rm ref}(C) \sum_{C'} p_{\rm ref}(C \to C') B(C') \\
&=& \sum_{C'} B(C') \sum_{C}  \pi_{\rm ref}(C) p_{\rm ref}(C \to C')  \\
&=& \sum_C \pi_{\rm ref}(C) B(C),
\eea
in which case
\beq
\theta_s = \sum_C \pi_{\rm ref}(C) \left\{ s B(C) + \ln \frac{R_{\rm ref}(C)}{R(C)} \right\}.
\eeq
For simple models for which states $C$ can be enumerated explicitly, the above expressions provide an analytic way of calculating the bound $I_0(a_s)= -\theta_s$. The forms \eq{int2} and \eq{int3} emphasize that only the typical occupancies $\pi_{\rm ref}(C)$ need be calculated; the form \eq{int3} emphasizes in addition that the prefactors of these occupancies can be derived from the object $\ln \sum_{C'} \e^{-s \alpha(C \to C')} W(C \to C')$. 

The rate function calculated using the reference-model method is not constrained to be convex: \eqq{ld2}, which is a direct evaluation of a probability, follows from simple arguments that do not depend on the form of the rate function in question. Convexity of the $s$-ensemble rate function, $I_s(a)$, is guaranteed by the structure of \eqq{leg}, which imposes a particular relationship between the two terms in that expression. Upon maximization, the quantity $a$ in the first term of \eq{leg} is given by the derivative of the second term, $-\partial \lambda(s)/\partial s$ (where this derivative exists). The same relationship does not exist between the terms of \eqq{ld2}, or between the terms of its bound, \eqq{bound}, $I_0(a_s) = -s a_s -q_s$. To see this, note from \eqq{int3} that
\beq
a_s = - \sum_C \pi_{\rm ref}(C) \partial_s \ln R_{\rm ref}(C),
\eeq
 and 
 \beq
 q_s = \sum_C \pi_{\rm ref}(C) \ln [R_{\rm ref}(C) /R(C)].
 \eeq
Thus
\beq
-\partial_s q_s = a_s  - \sum_C (\partial_s\pi_{\rm ref}(C)) \ln [R_{\rm ref}(C) /R(C)],
\eeq
and so $a_s$ and $q_s$ are not related in the way they would be if they were the elements of a Legendre transform (except in the special case of constant exit-rate ratio $R_{\rm ref}(C) /R(C)$). 

Thus the rate-function bound \eq{bound}, which is simple to calculate, can be non-convex and so can give information about coexistence of trajectories (see e.g. Fig. 4 of Ref.\c{klymko2017rare}).
     \begin{figure}[t] 
    \centering
    \includegraphics[width=\linewidth]{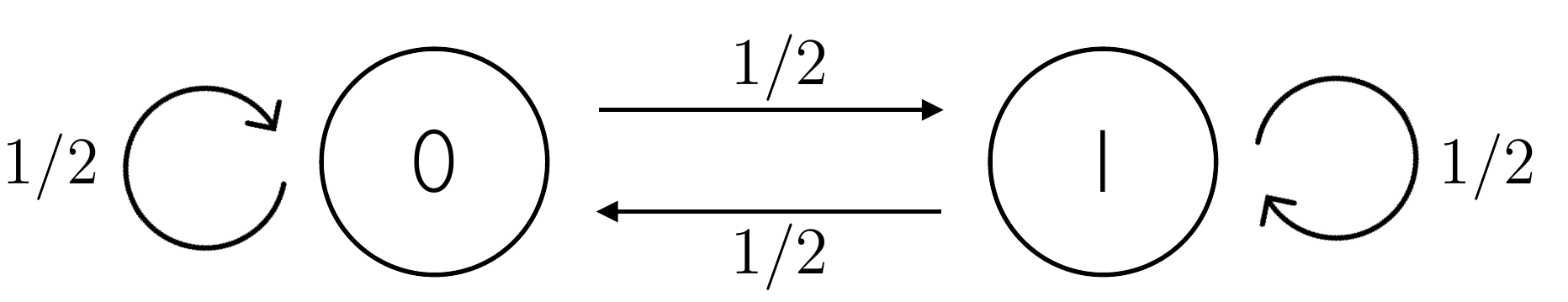} 
    \caption{The coin-toss model represented as a Markov process, with transition probabilities as indicated.}
    \label{fig_coin}
 \end{figure} 
 
\subsection{Example 1}
\label{sec_coin}

To illustrate the reference-model method we start with a simple example in which the $s$-ensemble and reference-model method are equivalent. Consider the unbiased coin-toss model, represented in \f{fig_coin} as a two-state Markov chain. The microstates $0$ and $1$ correspond to a tail or a head, respectively, and the probability of generating either is $1/2$. The master equation in the constant-$K$ ensemble, \eqq{me1}, has generator (matrix of transition probabilities)
\beq
\renewcommand{\arraystretch}{1.3}
{\bm W}=\left(
   \begin{array}{cc}
      \frac{1}{2} &  \frac{1}{2} \\
       \frac{1}{2} &  \frac{1}{2} \\
   \end{array}
   \right).
   \eeq
We define a dynamic observable $A$ that increases by $+1$ upon any move into microstate 1, and by $-1$ upon any move into microstate 0. The observable $a=A/K$ is then the total number of heads minus the total number of tails, divided by the number of coin tosses, and its rate function is\c{lewis1997introduction}
\beq
I_{\rm coin}(a) = \frac{1-a}{2} \ln \left(1-a\right)+\frac{1+a}{2}\ln \left( 1+a \right).
\eeq
The $s$-ensemble can be used to derive this result\c{touchette2009large}. The $s$-ensemble of the coin-toss model has generator
\beq
\renewcommand{\arraystretch}{1.3}
{\bm W}_s=\left(
   \begin{array}{cc}
      \frac{\e^s}{2} &  \frac{\e^s}{2} \\
       \frac{\e^{-s}}{2} &  \frac{\e^{-s}}{2} \\
   \end{array}
   \right),
   \eeq
whose principal eigenvalue is $\cosh s$. Then  $\lambda(s) = \ln \cosh s$, and \eqq{leg} yields the rate function
 \bea
 \label{here}
I_s(a) &=& \max_s(-s a - \ln \cosh s) \nonumber \\
&=& a \tanh^{-1}a - \ln \cosh \tanh^{-1} a \nonumber \\
&=& \frac{1-a}{2} \ln \left(1-a\right)+\frac{1+a}{2}\ln \left( 1+a \right), \hspace{0.8cm}
\eea
as expected. 
 
        \begin{figure*}[t] 
    \centering
    \includegraphics[width=0.7\linewidth]{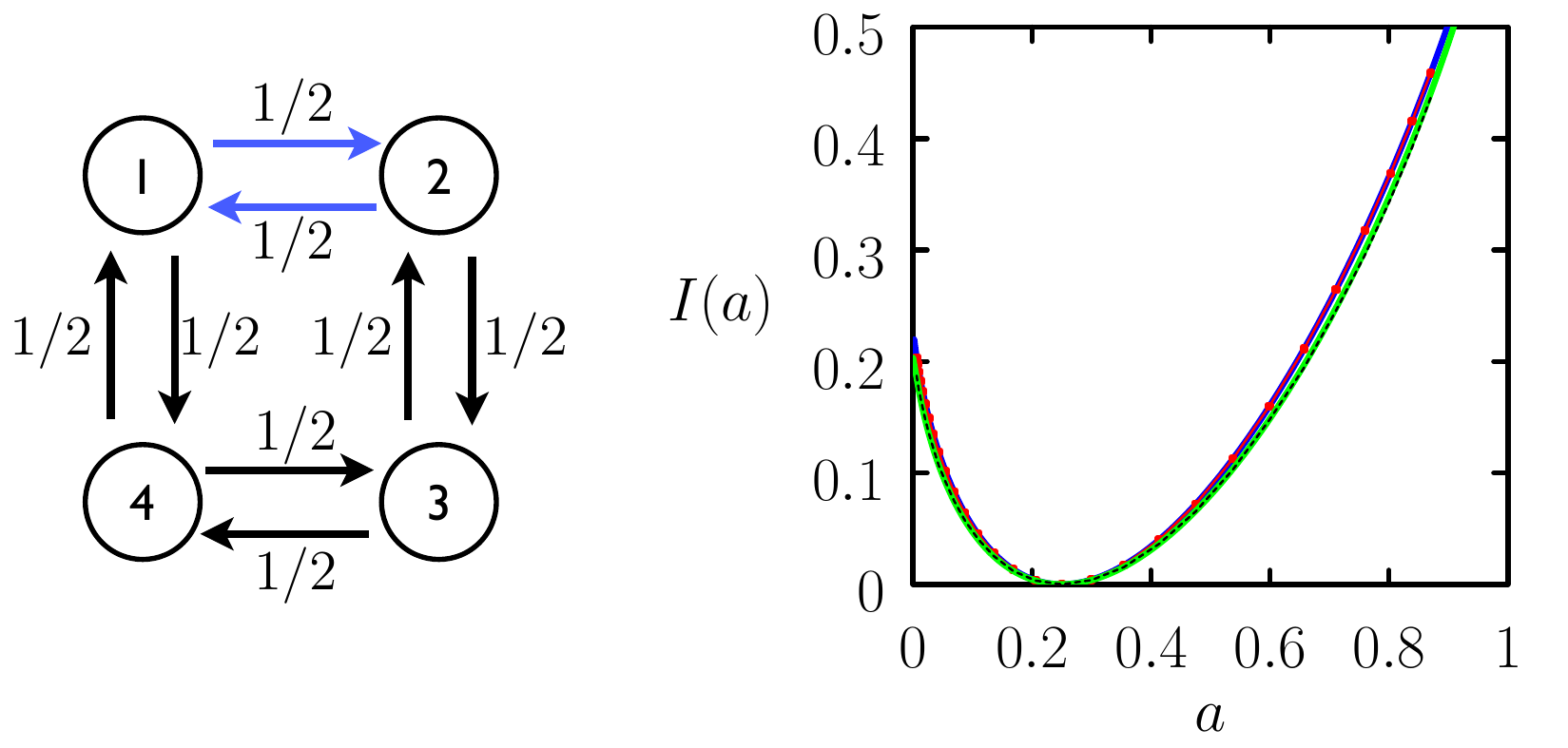} 
    \caption{For the 4-state model shown we compute the large-deviation rate function $I(a)$ associated with the order parameter $a=A/K$, where $A$ is incremented by 1 for every transition along the blue arrows in the course of a trajectory of $K$ steps. The green line is the result obtained using the $s$-ensemble. The blue and red lines on the right are upper bounds \eq{bound} on the rate function computed using the reference-model method. We keep track of explicit configuration changes during numerical simulations (red line, \eqq{int0}) or evaluate analytically \eqq{int2} $=$ \eqq{bound_an} (blue line). The black dashed line is obtained by evaluation of \eqq{rf_full}, which accounts for fluctuations of the weight for typical reference-model trajectories.}
    \label{4state}
 \end{figure*} 
Consider now the reference-model dynamics, which is generated by
\beq
\renewcommand{\arraystretch}{1.3}
{\bm W}_{\rm ref}=\left(
   \begin{array}{cc}
      \frac{\e^s}{2\cosh s} &  \frac{\e^s}{2 \cosh s} \\
       \frac{\e^{-s}}{2 \cosh s} &  \frac{\e^{-s}}{2 \cosh s} \\
   \end{array}
   \right).
   \eeq
This stochastic process can be simulated using standard methods\c{gillespie2005general}, in order to calculate \eqq{ld2} numerically\c{klymko2017rare}. For this problem the evaluation can also be done analytically. The steady-state measure of the reference model is $\pi(0) = \e^s/(2 \cosh s)$ and $\pi(1)=1-\pi(0)$, and the typical activity is $a_s = \sum_C \pi(C) \sum_{C'} p_{\rm ref}(C \to C') \alpha(C\to C') = -\tanh s$ (thus $a_s$ and $s$ have the relationship that $a$ and $s$ possess in \eqq{here}). The exit-rate ratio is always $q_s = \ln(R_{\rm ref}(C)/R(C)) = \cosh s$, and so \eqq{ld2} is
   \bea
   I(a_s) &=& a_s \tanh^{-1} a_s - \ln \cosh \tanh^{-1} a_s \nonumber \\
   &=& \frac{1-a_s}{2} \ln \left(1-a_s\right)+\frac{1+a_s}{2}\ln \left( 1+a_s \right), \hspace{0.8cm}
   \eea
giving us one point on the rate-function curve $I_{\rm coin}(a)$. Repeating the method for a range of values of $s$ (which here amounts simply to replacing $a_s \to a$) allows us to reconstruct the full rate-function curve.

\subsection{Example 2}

A second example illustrates a case in which the $s$-ensemble and reference-model ensemble are different (but both can be used to obtain $\rho(a,K)$). Consider the 4-state model shown in \f{4state}, with each rate set to unity (indicated are the transition probabilities, all $1/2$). Define the activity $A$ as the number of configuration changes from state 1 to 2 or state 2 to 1 in the course of a simulation of $K$ steps. By symmetry of the network the mean number of such configuration changes will be $A=K/4$, but some trajectories will exhibit more changes, and some less. We can use the $s$-ensemble and the reference-model method to calculate the logarithmic probability distribution $I(a) = -K^{-1} \ln \rho(a,K)$ associated with the quantity $a=A/K$.

The generator of the original model is 
\beq
\renewcommand{\arraystretch}{1.3}
{\bm W}=\left(
   \begin{array}{cccc}
     0 & \frac{1}{2} & 0 &\frac{1}{2} \\
       \frac{1}{2} & 0 &\frac{1}{2} & 0\\
            0 & \frac{1}{2} & 0 &\frac{1}{2} \\
                   \frac{1}{2} & 0 &\frac{1}{2} & 0
   \end{array}
   \right),
   \eeq
and that of the $s$-ensemble is
\beq
\renewcommand{\arraystretch}{1.3}
{\bm W}_s=\left(
   \begin{array}{cccc}
     0 & \frac{\e^{-s}}{2} & 0 &\frac{1}{2} \\
       \frac{\e^{-s}}{2} & 0 &\frac{1}{2} & 0\\
            0 & \frac{1}{2} & 0 &\frac{1}{2} \\
                   \frac{1}{2} & 0 &\frac{1}{2} & 0
   \end{array}
   \right).
   \eeq
 The principal eigenvalue of ${\bm W}_s$ is $e^{\lambda(s)} = \frac{1}{4} \left(\e^{-s} \sqrt{-2 \e^s+5 \e^{2 s}+1}+\e^{-s}+1\right)$, from which, using \eqq{leg}, $I(a)$ can be calculated numerically. The result is the green line in \f{4state}.
  \begin{figure*}[t] 
    \centering
    \includegraphics[width=0.8\linewidth]{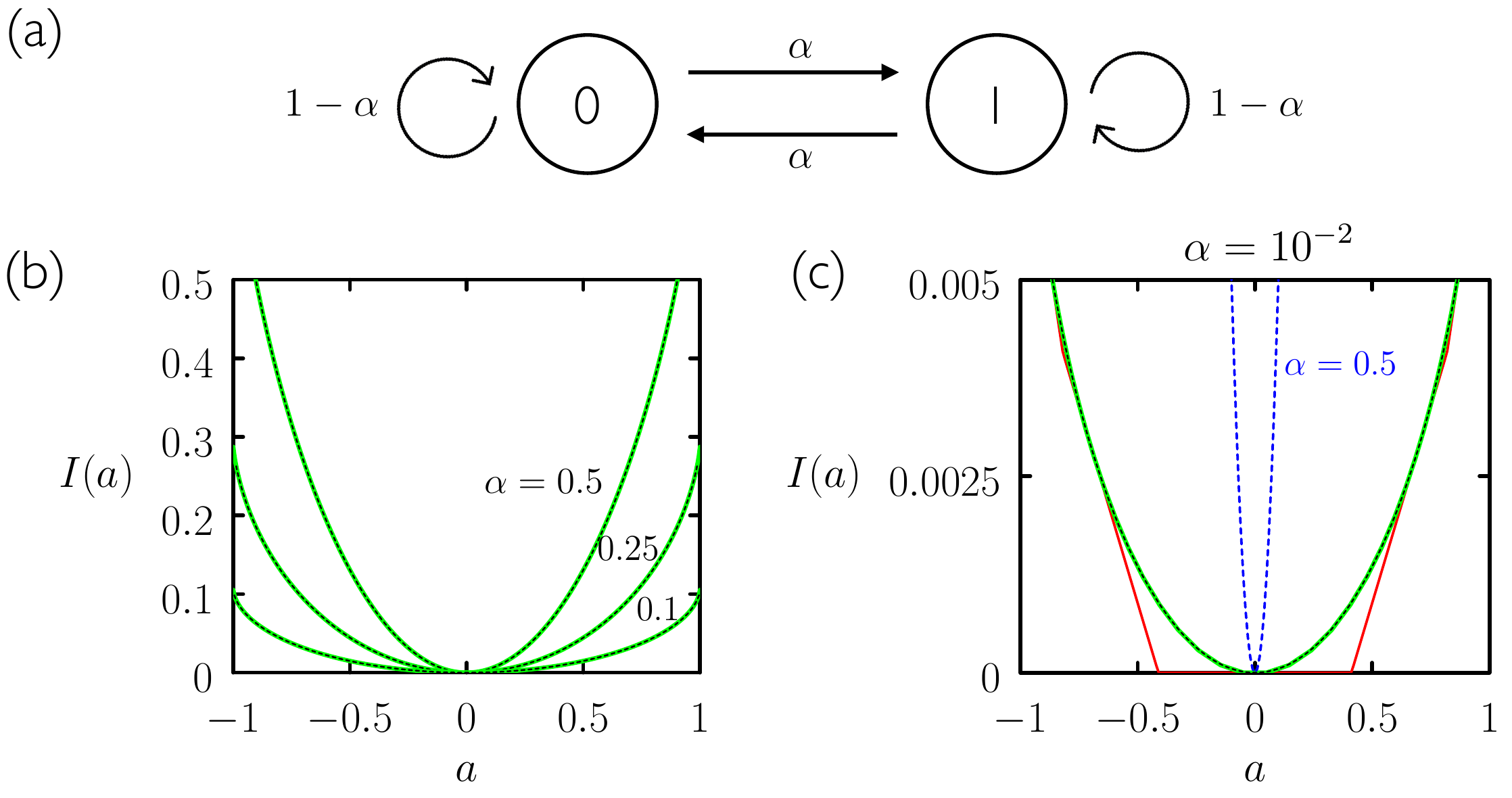} 
    \caption{(a) Minimal model of a one-dimensional active particle\c{touchette2009large}. (b) For the values of $\alpha$ shown, rate functions obtained from the $s$-ensemble (green) and reference-model method (black dashed lines) agree. (c) For small values of $\alpha$ the rate function is very broad (red, green, and black lines, $\alpha = 10^{-2}$; the blue line shows the case $\alpha=1/2$ for comparison). Here the rate function can be obtained from the $s$-ensemble via Legendre transform (green line) only if a very fine numerical grid of $s$ is used; if not, the rate function displays spurious flat portions (red line). The reference-model method (black line) evaluates $I(a)$ directly, and does not require closely-spaced values of $s$.}
    \label{fig_walker}
 \end{figure*} 
 
 The reference-model generator is
 \beq
\renewcommand{\arraystretch}{1.3}
{\bm W}_{\rm ref}=\left(
   \begin{array}{cccc}
     0 & \frac{\e^{-s}}{1+\e^{-s}} & 0 &\frac{1}{2} \\
      \frac{\e^{-s}}{1+\e^{-s}}  & 0 &\frac{1}{2} & 0\\
            0 & \frac{1}{1+\e^{-s}}  & 0 &\frac{1}{2} \\
                   \frac{1}{1+\e^{-s}} & 0 &\frac{1}{2} & 0
   \end{array}
   \right),
   \eeq
which can be simulated using standard stochastic methods\c{gillespie2005general}. For a range of values of $s$ (from $-3$ to $3$, in intervals of 0.25) we carried out a single simulation of the reference model of $K=5 \times 10^8$ configuration changes. We computed the rate-function bound \eq{bound} by a) keeping track of explicit configuration changes, as in \eqq{int0}, and b) by evaluating \eqq{int2}. To evaluate the latter, note that the steady-state measure is
\beq
\pi_{\rm ref}(1) = \pi_{\rm ref}(2) =  \frac{1+\e^{-s}}{2(3+\e^{-s})}.
\eeq
The typical activity of the reference model is then $a_s = (\pi_{\rm ref}(1)+ \pi_{\rm ref}(2)) \times \e^{-s}/(1+\e^{-s})$, from which we obtain $s=\ln[(1/a_s-1)/3]$. The relationship between $a_s$ and $s$ is different to the relationship $a = -\partial_s \lambda(s)$ obtained from the $s$-ensemble: here the two methods produce different typical activity for given $s$. 

From \eqq{int2} we have
\beq
\label{bound_an}
I_0(a) = -\theta_s(a) = -a \ln \frac{1-a}{3a} - \frac{1+2a}{3}  \ln\frac{1 + 2 a}{2 - 2 a}.
\eeq
As shown in \f{4state}, both ways of calculating $I_0(a)$, analytically using \eqq{bound_an} and numerically using \eq{int0}, the blue and red lines in the figure, give the same result. 

The full rate function can be estimated by evaluating numerically the integral in \eqq{ld2}. To do so we took each trajectory of $5 \times 10^8$ steps and split it into $5 \times 10^3$ pieces of length $K'=10^5$. Of that set we retained only those pieces with extensive order parameter $A' = K' a_s$ (it is necessary to explicitly enforce the delta-function constraint of \eqq{summ} for short trajectories). The resulting distribution of $q$, $P(q|a_s)$, is Gaussian with variance $\sigma_s^2$. In this case \eqq{ld2} reads
\beq
\label{rf_full}
I(a_s) = -s a_s-q_s- K \sigma_s^2/2.
\eeq
Evaluation of \eqq{rf_full} gives the dotted black line in \f{4state}, consistent with the result calculated via the $s$-ensemble.

\subsection{Example 3}
   
 A third example illustrates some numerical difficulties associated with obtaining a rate function via Legendre transform, even when the rate function is convex. Consider the two-state Markov chain of example IV.4 of\c{touchette2009large} (see also\c{oono1989large}), shown in \f{fig_walker}(a). We shall construct a dynamic observable $A$ by associating a value of $+1$ with any move into state 1, and $-1$ with any move into state 0. In this case we can consider the model to be a minimal representation of a one-dimensional active particle (see e.g.\c{pietzonka2016extreme,thompson2011lattice}), with 0 being a left-pointing, left-moving state, and 1 being a right-pointing, right-moving state. At each step the particle flips direction with probability $\alpha$ (and takes a step in the new direction), or else retains its current orientation and takes a step in the corresponding direction. $a=A/K$ is then the position of the particle after $K$ steps. 
     \begin{figure}[t] 
    \centering
    \includegraphics[width=\linewidth]{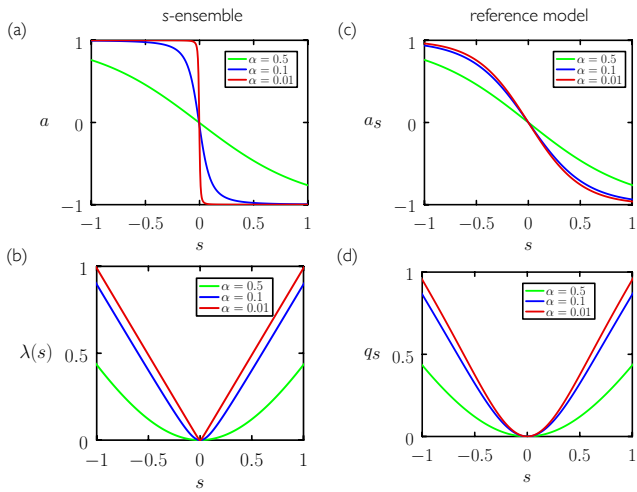} 
    \caption{The activity $a= -\partial_s \lambda(s)$ (panel (a)) and scaled cumulant-generating function $\lambda(s)$ (b) in the $s$-ensemble are constrained to vary sharply whenever the rate function $I(a)$ of the original model (shown in \f{fig_walker}) has small gradient. In this case this is so near $a=0$ as $\alpha$ becomes small. As a result, fine sampling near $s=0$ is required to allow recovery of $I(a)$ by Legendre transform of $\lambda(s)$; see \f{fig_walker}(b). By contrast, in the reference-model method the quantities $a_s$ (c) and $q_s$ (d) are not constrained to vary sharply when $I(a)$ has small gradient. Their combination $-s a_s -q_s$ yields the piece $I(a_s)$ of the rate function $I(a)$.}
    \label{fig_aux}
 \end{figure} 
 
 For $\alpha=1/2$ we have the random walker (or unbiased coin-toss) of \s{sec_coin}. For $\alpha=0$ the two states cannot interconvert (the Markov chain is reducible), and there exist two rate functions (which vanish for $a=1$ or $a=-1$, and are infinite otherwise). For $0 < \alpha <1/2$ we have an `active walker' that moves persistently right or left between flips. In physical terms, fluctuations of $A/K$ for this model are related to fluctuations of magnetization $M/N$ in the one-dimensional Ising model in zero magnetic field, where $N$ is system size. Flips between runs $\cdots {\rm LLLRRRRLLLL} \cdots$ occur independently with probability $\alpha$, much as domain walls between Ising spins $\cdots ---++++----\cdots$ occur independently with probability $(1+\e^{\beta J})^{-1}$, where $J$ is the Ising coupling. For nonzero $\alpha$ (finite $J$) we expect domain lengths to be exponentially distributed with mean $\alpha^{-1}$ or $1+\e^{\beta J}$; for there to be no bias between L and R or between $-$ and $+$; and for the mean values of $A/K$ and $M/N$ in the limits of large $K$ and large $N$ to be zero. Thus the rate function $I(a)$ should be convex, symmetric, and have a (unique) minimum at zero. 
 
We first calculate the rate function using the $s$-ensemble, following Ref.\c{touchette2009large}. The generator
 \beq
{\bm W}= \left(
\begin{array}{cc}
 1-\alpha  & \alpha  \\
 \alpha  & 1-\alpha 
\end{array}
\right)
 \eeq
 becomes, in the $s$-ensemble,
 \beq
{\bm W}_s= \left(
\begin{array}{cc}
 (1-\alpha )\e^s & \alpha  \e^s \\
 \alpha  \e^{-s} & (1-\alpha ) \e^{-s}
\end{array}
\right).
 \eeq
 From this we extract $\lambda(s)$ in the usual way\c{touchette2009large}, and calculate $I(a)$ via Legendre transform. Some results for three values of $\alpha$ are shown in \f{fig_walker}(b) (green lines).

We next calculate $I(a)$ using the reference-model method. Its generator is 
 \beq
 \renewcommand{\arraystretch}{1.5}
 {\bm W}_{\rm ref}=\left(
\begin{array}{cc}
 \frac{(1-\alpha ) \e^s}{ (1-\alpha ) \e^s+\alpha  \e^{-s}} &
   \frac{\alpha  \e^s}{ (1-\alpha ) \e^{-s}+\alpha  \e^s} \\
 \frac{\alpha  \e^{-s}}{ (1-\alpha ) \e^s+\alpha  \e^{-s}} &
   \frac{(1-\alpha ) \e^{-s}}{ (1-\alpha )\e^{-s}+\alpha  \e^s}
\end{array}
\right).
 \eeq
 Calculation of $I(a)$ can be done analytically, because the Jensen bound \eq{bound} is exact for this model. This is so because each value of activity $A$ is associated with a unique path weight: every move into state 1 (and only state 1) generates activity +1, and every move out of the state generates a piece $\ln (R_{\rm ref}(1)/R(1))$ of the path weight. Similarly for state 0. By conservation of probability, every move into a state must be balanced by a move out of the state, and so the path weight takes a unique value for each value of $A$. Solving as before for the stationary measure $\pi_{\rm ref}(C)$, we evaluate the terms in \eqq{int2} and write the rate function in the parametric form $(a_s(s),I(s))$; we get
 \beq
 \label{a1}
 a_s(s) = \frac{(\alpha -1) \sinh (2 s)}{\alpha -(\alpha -1) \cosh (2 s)}
 \eeq
 and
  \begin{widetext}
 \bea
 \label{a2}
I(s)&=& \frac{(\alpha -1) \left(\e^{4 s}-1\right) s+\left(\alpha  \left(\e^{2 s}-1\right)+1\right) \ln ((2 \alpha -1) \sinh
   s+\cosh s)}{\alpha  \left(\e^{2 s}-1\right)^2-\e^{4 s}-1} \nonumber \\
   &+&\frac{\e^{3 s} (-2 \alpha  \sinh s+\sinh s+ \cosh s) \ln (-2 \alpha  \sinh s+\sinh s+\cosh
   s)}{\alpha  \left(\e^{2 s}-1\right)^2-\e^{4 s}-1}.
 \eea
 \end{widetext}
 In \f{fig_walker}(b) we show $I(a)$ obtained from the reference-model method for three values of $\alpha$ (black dotted lines). These results agree with those obtained using the $s$-ensemble.
 
 For smaller values of $\alpha$ the Legendre transform encounters numerical problems. In \f{fig_walker}(c) we consider the case $\alpha=10^{-2}$ (the blue line shows the case $\alpha=1/2$ for comparison). The rate function, calculated via the reference-model method, is shown in black. The same result can be obtained using the $s$-ensemble via numerical Legendre transform, but only if many closely-spaced values of $s$ are used to calculate \eqq{leg}: the associated scaled cumulant-generating function is sharply kinked, and its numerical differentiation requires fine intervals of $s$. In \f{fig_coin}(c) the green line was calculated by numerical Legendre transform using intervals of $s$ of $5 \times 10^{-4}$, and this agrees with the reference-model result. The red line was calculated by Legendre transform using coarser intervals of spacing $5 \times 10^{-3}$, and the result has spurious flat portions. It is clear that this result is unphysical: the mean value of $a$ of the original model is 0, and so the true rate function must have a unique minimum at zero. The smaller the value of $\alpha$, the finer the mesh required by the Legendre transform.
 
 In this case the rate function is `simple', i.e. it is convex and quadratic (for $\alpha$ small we can expand \eq{a1} and \eq{a2} about $s=0$ to obtain $I(a) \approx \alpha a^2/2$) but its calculation via Legendre transform requires fine sampling of values of $s$. This is so for geometrical reasons. In the $s$-ensemble the calculated quantity $\lambda(s)$ is the Legendre transform of the rate function $I(a)$. As a result (see Fig. 4 of\c{touchette2009large}), $s(a)$ is given locally by (minus) the slope of the rate function $I(a)$. Thus $a$ varies sharply with $s$ wherever the rate function has small curvature. In \f{fig_aux}(a) we show $a =- \partial_s \lambda(s)$ in the $s$-ensemble, for the model of \f{fig_walker}; $a$ changes sharply near $s=0$ as $\alpha$ becomes small (here $a \approx -s/\alpha$ near $s=0$). The corresponding scaled cumulant-generating function (b) becomes sharply kinked, and fine sampling near $s=0$ is needed to convert $\lambda(s)$ into $I(a)$. 
 
By contrast, the key quantities $a_s$ and $q_s$ of the reference-model method, panels (c) and (d) of \f{fig_aux}, are not constrained to change sharply where original model's rate function $I(a)$ is slowly varying. The reference-model rate functions shown in \f{fig_walker} are evaluated directly, by taking the combination $-s a_s - q_s$. Each point on the function can be evaluated independently, and the grid of $s$ can be as coarse or as fine as desired. The evaluation works when the rate function has very small curvature or is non-convex (see e.g. Fig. 4 of\c{klymko2017rare}). In the latter case the rate function cannot be recovered by Legendre transform\c{touchette2009large}.

In physical terms the results of this section relate to the probability distribution $P(a)$ of the position $a$ of an active particle, with rotation rate $\alpha$, in discrete space. This distribution adopts a large-deviation form $P(a) \sim \e^{-K I(a)}$ in the limit of a large number of steps $K$, with $I(a)$ given by Equations~\eq{a1} and~\eq{a2}, shown in \f{fig_walker}. For small $\alpha$, fluctuations of $a$ become large in the sense of having a large variance, but are to leading order Gaussian: $I(a)\approx \alpha a^2/2$, for $a, \alpha$ small. This distribution can be calculated using the $s$-ensemble, whose key quantities $\lambda(s)$ and $a(s)$ (see \f{fig_aux}(a,b)) show a sharp jump when the rate function has small gradient. This is so for the geometrical reasons shown in Fig. 4 of\c{touchette2009large}. The distribution $P(a)$ can also be calculated using the reference-model method, whose key quantities (see \f{fig_aux}(c,d)) are not required to change sharply. Indeed, here the original model is similar to the 1D Ising model at nonzero temperature, for the reasons discussed above, and the set of reference models is similar to the 1D Ising model in a field $h \propto s$. Neither shows a sharp transition. 

\section{Sampling nonequilibrium trajectories of a lattice model of active matter.}
\label{sec_active}

\begin{figure}[h] 
    \centering
    \includegraphics[width= \linewidth]{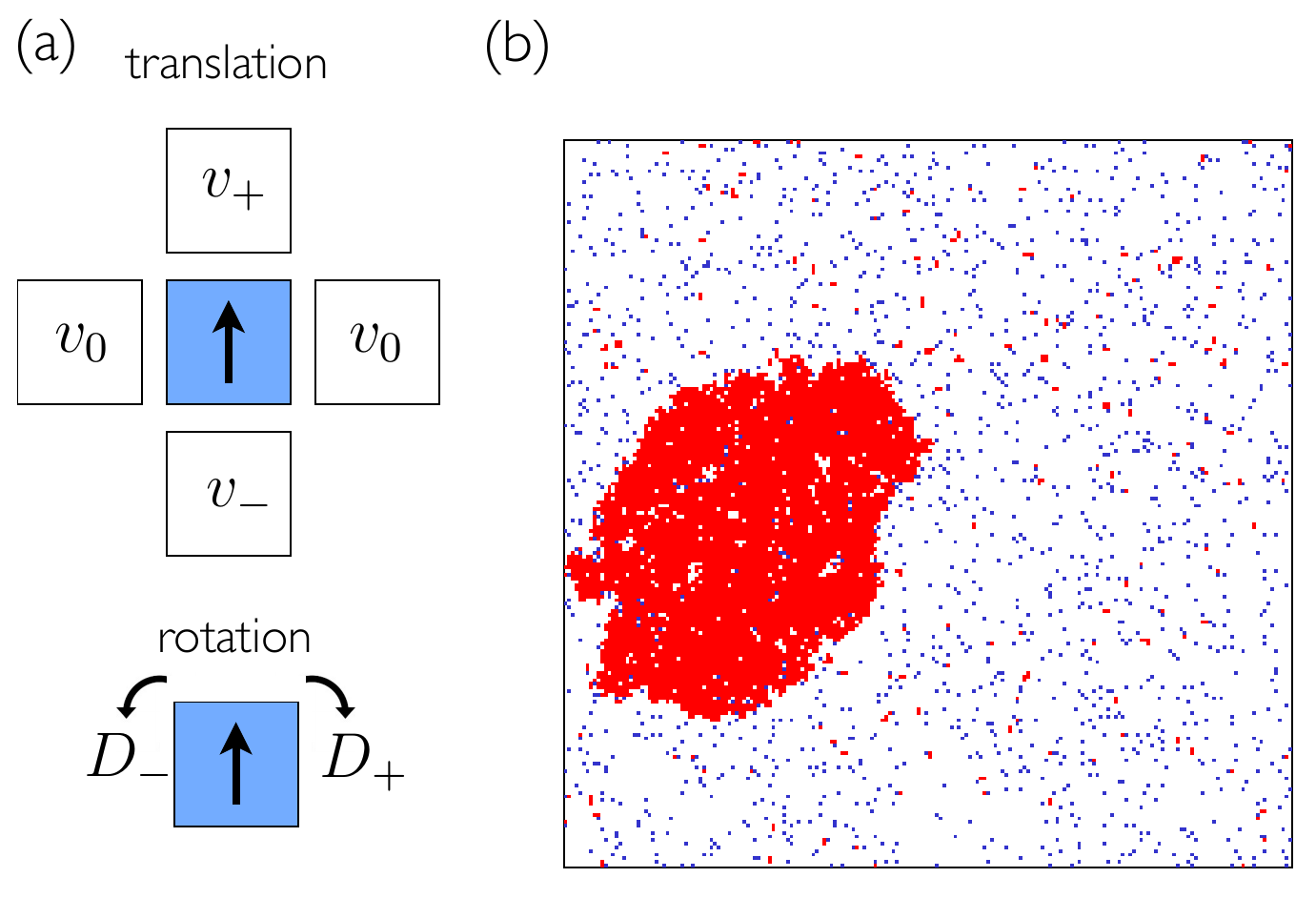} 
    \caption{(a) Rates of motion for a model of lattice-based active particles\c{lattice1}. (b) A typical phase-separated configuration of collections of volume-excluding particles of this kind, produced using a continuous-time Monte Carlo algorithm\c{gillespie2005general}, starting from well-mixed conditions. The packing fraction is $\phi=1/5$; the rates are $v_+ = 16$, $v_-=v_0=1$, $D_+=D_-=1/10$. The lattice size is $200^2$, and periodic boundaries are applied in each direction. Particles are colored red if they point toward a nearest-neighbor particle, and blue otherwise.}
    \label{fig1}
 \end{figure} 
  \begin{figure*}[t] 
    \centering
\includegraphics[width=\linewidth]{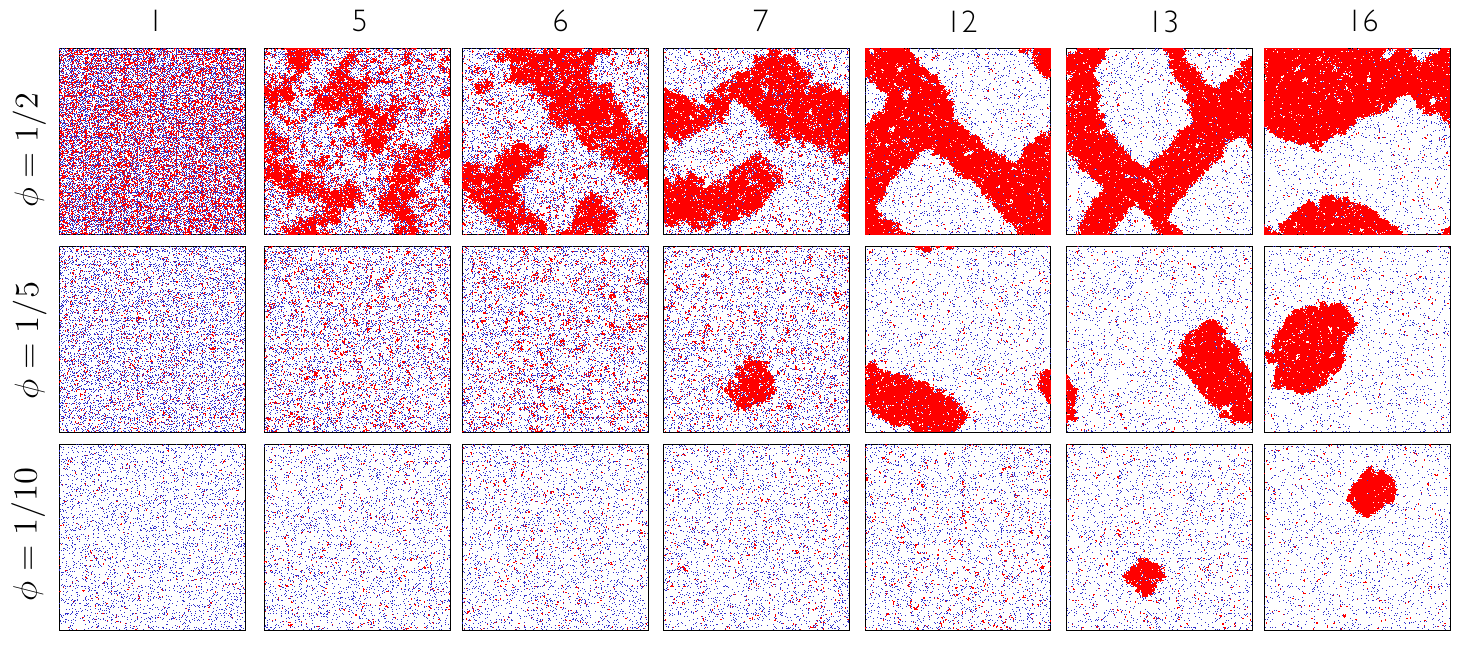} 
    \caption{Active lattice-based particles undergo phase separation at a density-dependent value of their motility. Rows are labeled by the packing fraction, $\phi$, and columns are labeled by the rate of forward motion, $v_+$. The other rates are as \f{fig1}, giving a P\'eclet number ${\rm Pe} = 5(v_+-1)$\c{lattice1}. The lattice size is $200^2$.}
    \label{fig2}
 \end{figure*} 
 
 For models with larger state space, matrix diagonalization becomes impractical, and other methods are required to determine the $s$-ensemble\c{touchette2009large,maes2008canonical,giardina2011simulating,lecomte2007numerical,jack2015effective,nemoto2014computation,nemoto2016population}. The reference-model method is implemented in the same way, however, as we now illustrate using the lattice model of active particles introduced in Ref.\c{lattice1}. 
 
Active particles propel themselves by consuming energy, and move by diffusion and by drift in a direction that fluctuates\c{tailleur2008statistical,weitz2015self,baskaran2009statistical,peruani2012collective,barre2015motility,peruani2011traffic,ramaswamy2010mechanics, Romanczuk_2012,redner2013structure,marchetti2013hydrodynamics,buttinoni2013dynamical, stenhammar2013continuum, Yeomans_2014, cates2015motility, Menzel_2015, Bechinger_2016}. Active particles form clusters and undergo phase separation even in the absence of interparticle attractions\c{ramaswamy2010mechanics,tailleur2008statistical,Fily_2012,redner2013structure,Speck_2014,cates2015motility}. This phenomenon, called motility-induced phase separation, can be reproduced on the computer using simple model particles\c{Fily_2012,redner2013structure,Speck_2014,thompson2011lattice} (motility-induced phase separation also occurs in the presence of velocity alignment\c{solon2015flocking,solon2013revisiting,peruani2011traffic,PhysRevE.89.012718}). Associated with this phenomenon are various kinds of large fluctuations\c{peruani2010cluster,Fily_2012,peruani2012collective}.

With an eye to studying such fluctuations in detail we introduced in Ref.\c{lattice1} the following lattice model of active matter. Consider particles that move in isolation with the rates shown in \f{fig1}(a). Particles possess both positions and orientations, and move on a two-dimensional square lattice. An isolated particle moves forward with rate $v_+$, backward with rate $v_-$, to either side with rate $v_0$, and rotates $\pi/2$ with rate $D_+$ or $D_-$ (here we set $D_+=D_-\equiv D_{\rm rot}$). An on-lattice particle of this nature moves in a manner similar to an off-lattice Brownian particle\c{lattice1}. As expected from the behavior of off-lattice particles\c{Fily_2012,redner2013structure,Speck_2014,thompson2011lattice}, on-lattice active particles undergo motility-induced phase separation: see \f{fig1}(b) and \f{fig2}. We used $N$ hard particles on a periodically-replicated lattice of size $L^2$ (for a packing fraction $\phi=N/L^2$), starting from disordered configurations, and propagated dynamics using a continuous-time Monte Carlo algorithm\c{gillespie2005general}. At each step of the simulation the system moves from the current configuration $C$ to a new configuration $C'$ with probability
\beq
p(C \to C') = \frac{W(C \to C')}{\sum_{C'} W(C \to C')},
\eeq
where the rates $W$ for each possible process are taken from \f{fig1}. Typical trajectories of this model lead to phase separation at a density-dependent value of the P\'eclet number\c{lattice1,redner2013structure}.
\begin{figure*}[] 
    \centering
   \includegraphics[width=0.9\linewidth]{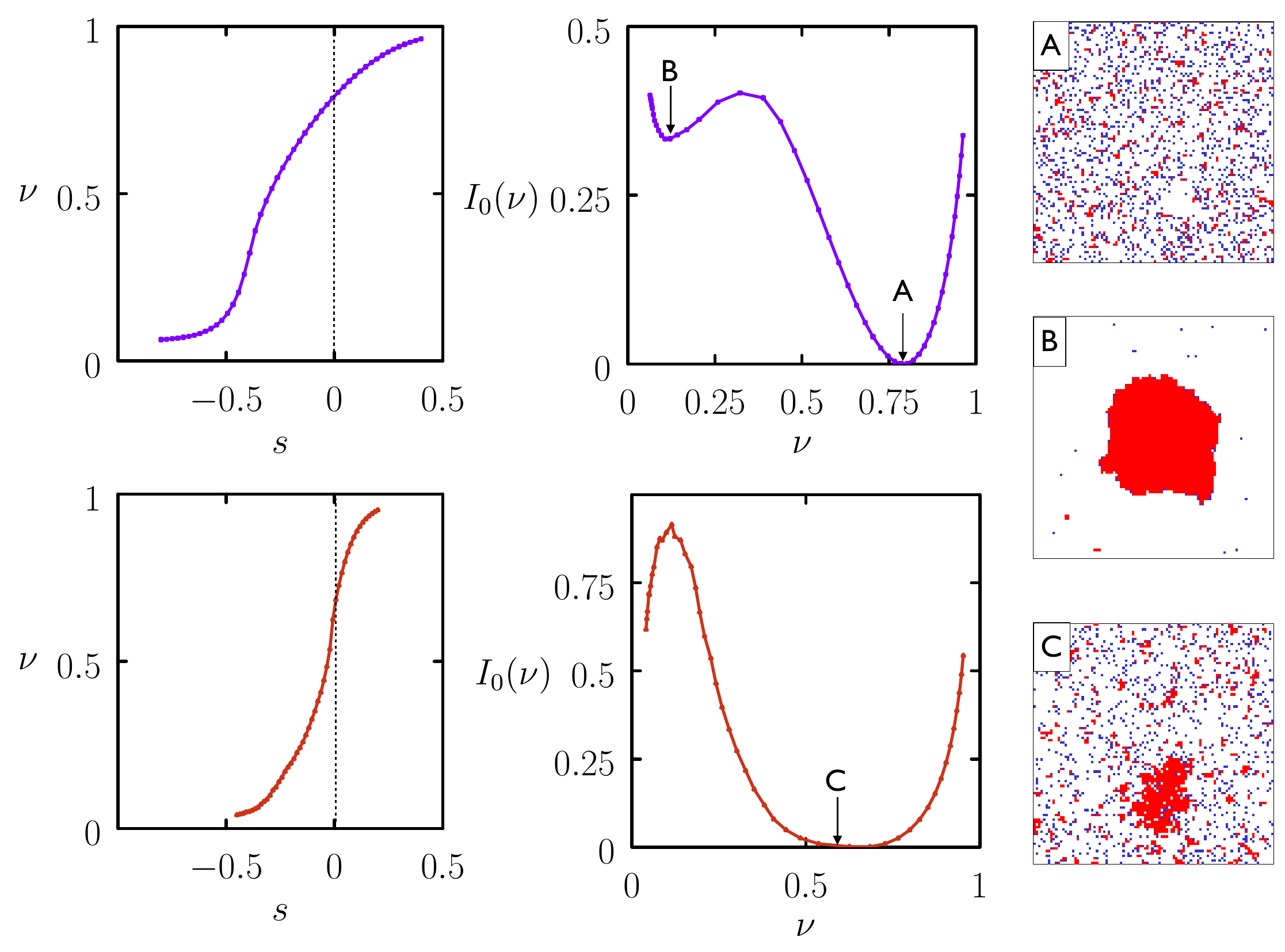} 
    \caption{Activity $\nu$ as a function of $s$ (left graphs) and rate-function bound $I_0(\nu)$ as a function of $\nu$ (right graphs) from the reference-model method applied to the active lattice model. Packing fraction is $\phi=1/5$ and drift rate $v_+=2$ (blue lines, top) or $v_+=7$ (red lines, bottom). Lattice size is $100^2$. Trajectories prepared at $s=0$ are typical of the {\em original} model, while trajectories prepared at $s\neq0$ are rare in the original model; how rare can be understood from the right-hand graphs. For $v_+=2$ we are far from the `typical' phase transition (see \f{fig2}), and the crossover from active to inactive trajectories occurs for $s \neq 0$. For $v_+=7$, close to the `typical' transition, the crossover occurs near $s=0$. Shown right are typical and rare configurations of the model.}
    \label{s_pic}
 \end{figure*}
 
To quantify the rare behavior of this model we chose a dynamic order parameter $a$ motivated by studies of glasses. There, authors sometimes choose to count the number of events that occur in a particular time, with the mean time taken to leave configuration $C$ being $1/R(C)$. This choice allows the identification of phase transitions out of equilibrium\c{garrahan2009first,hedges2009dynamic,budini2014fluctuating}. Here we choose to measure the values of the escape rates $R(C)$ of states explored by a fixed number of configuration changes. With this choice the bias $\alpha(C\to C')=B(C')$, where $B(C') = R(C')$ is the escape rate from the state to which the model is moving, and the transition probabilities of the reference model are
\beq
p_{\rm ref}(C \to C') = \frac{\e^{-s[B(C')-B(C)]}W(C \to C')}{\sum_{C'} \e^{-s[B(C')-B(C)]} W(C \to C')}.
\eeq
The factors $\e^{s B(C)}$ in numerator and denominator have been introduced so that numbers appearing in the arguments of exponentials are not too large. 

The coding overhead for the reference model is only slightly greater than that of the original model. One needs to keep track of the escape rate $R(C')$ of each configuration $C'$ accessible from the current configuration (rather than just the escape rate of the current configuration), but this can be done efficiently, in the course of the simulation, by noting which particles can touch the moving particle, before and after its move, one move into the future. 

The order parameter against which dynamics is conditioned is $ a= A / K = K^{-1} \sum_{k=0}^{K-1} B(C_{k+1})$, the mean relaxation rate of configurations comprising the trajectory, which we report in the form of a scaled `activity', $\nu \equiv a/(\Sigma N)$. Here $N$ is the number of particles and $\Sigma \equiv v_++2v_0+v_-+2D_{\rm rot}$ is the sum of rates of processes of isolated particles. We used \eqq{int0} to calculate the rate-function bound, which in this case reads
\begin{eqnarray}
\label{rf2}
I_0(a_s) & = & -s K^{-1}\sum_{k=0}^{K-1} [B(C_{k+1})-B(C_k)]  \\
& & - K^{-1}\sum_{k=0}^{K-1} \ln \frac{\sum_{C'}{\rm e}^{-s [B(C')-B(C_k)]} W(C_k \to C')}{\sum_{C'}W(C_k \to C')}. \nonumber
\end{eqnarray}
The first term on the right-hand side of \eq{rf2} becomes negligible for large $K$. We evaluated this expression for trajectories of about $K=10^8$ steps (the bound looks no different trajectories of $K \sim 10^7$ steps are used). 

Results derived from these simulations are shown in \f{s_pic}, for packing fraction $\phi=1/5$. Here the rate-function bound (which is not convex) shows that the typical activity of the model (where $I_0(\nu)= 0$) shifts from large $\nu$ to small $\nu$ as the drift rate $v_+$ of the active particles increases, because typical trajectories exhibit clustering (see \f{fig2}) and (hence) reduced dynamism. In addition, trajectories show an active-to-less active crossover as a function of the bias parameter $s$. For model parameters far from the `typical' nonequilibrium phase transition (which occurs near $v_+ \approx 7$ for $\phi=1/5$) this crossover is seen in the ensemble of rare trajectories (for which $s \neq 0$), while for model parameters near the `typical' transition the crossover occurs in the ensemble of typical trajectories, for which $s \approx 0$. This behavior resembles transitions observed in models of glasses\c{hedges2009dynamic} or proteins\c{mey2014rare} using the $s$-ensemble.

\section{Conclusions}
\label{sec_conc}

We have described a method to sample the rare fluctuations of discrete-time Markov chains, and have applied it to the lattice-based model of active matter introduced in\c{lattice1}. The method is inspired by the $s$-ensemble method but is distinct: we use a probability-conserving reference model to calculate directly the likelihood that a model displays a particular value of a dynamic order parameter $a$. As a result, the method can be used to calculate non-convex rate functions, so giving information about trajectory coexistence near nonequilibrium phase transitions. The method is straightforward to implement for models for which continuous-time Monte Carlo dynamics can be carried out. Take the rates of the original model $W(C \to C')$, change them to $\e^{-s \alpha(C \to C')} W(C \to C')$, run simulations and evaluate the quantities needed for equation \eq{ld2} (repeat for a range of values of $s$). In general, as noted in Ref.\c{rohwer2015convergence}, high-precision evaluation of integrals of the form appearing in \eqq{ld2} is demanding. For several models we have studied, however, calculation of the variance of $q$ appears to give a good approximation of the rate function $I(a)$ (because fluctuations of $q$ are often approximately Gaussian), while calculation of the mean of $q$ -- which is straightforward -- gives via \eqq{bound} a physically meaningful, and potentially non-convex, rate-function bound. Thus the method provides a simple way of obtaining non-trivial information about a model and its phase transitions away from equilibrium, and so provides a complement to existing trajectory-sampling methods\c{ruelle2004thermodynamic,garrahan2007dynamical,Lecomte2007,garrahan2009first,hedges2009dynamic,bucklew1990large,chetrite2015variational,touchette2009large,giardina2006direct,maes2008canonical,giardina2011simulating}. 

We end with a note about time. The reference-model method operates within the constant-event-number ensemble (the discrete-time Markov chain), because placing no restriction upon jump times makes the sampling procedure easier (see Appendix B of\c{klymko2017rare}). However, the procedure used to move through configuration space is the same as that used in a continuous-time Monte Carlo scheme (except that no random number is drawn to compute the time of escape from a given state), and all temporal information can be recovered after the fact. The escape time $t_k$ from configuration $C_k$ is an exponentially distributed random variable with mean $1/R(C_k)$, and so the total time of a trajectory, $T=\sum_{k=0}^{K-1} t_k$, is a stochastic variable with a known distribution $P(T)$\c{jasiulewicz2003convolutions}. This distribution reduces to the Erlang distribution if all $R(C_k)$ are identical, and to the hypoexponential distribution if all $R(C_k)$ are distinct\c{buchholz2014input}. Thus the complete distribution of trajectory times, for fixed number of events and specified activity $a$, can be recovered from the constant-event-number ensemble without additional simulation.

\vspace{0.4cm}

\begin{acknowledgements}
I thank Hugo Touchette, Juan Garrahan, and Katherine Klymko for discussions. This work was performed at the Molecular Foundry, Lawrence Berkeley National Laboratory, supported by the Office of Science, Office of Basic Energy Sciences, of the U.S. Department of Energy under Contract No. DE-AC02--05CH11231. 
\end{acknowledgements}
%\bibliography{bib_v2}  

\begin{thebibliography}{61}%
\makeatletter
\providecommand \@ifxundefined [1]{%
 \@ifx{#1\undefined}
}%
\providecommand \@ifnum [1]{%
 \ifnum #1\expandafter \@firstoftwo
 \else \expandafter \@secondoftwo
 \fi
}%
\providecommand \@ifx [1]{%
 \ifx #1\expandafter \@firstoftwo
 \else \expandafter \@secondoftwo
 \fi
}%
\providecommand \natexlab [1]{#1}%
\providecommand \enquote  [1]{``#1''}%
\providecommand \bibnamefont  [1]{#1}%
\providecommand \bibfnamefont [1]{#1}%
\providecommand \citenamefont [1]{#1}%
\providecommand \href@noop [0]{\@secondoftwo}%
\providecommand \href [0]{\begingroup \@sanitize@url \@href}%
\providecommand \@href[1]{\@@startlink{#1}\@@href}%
\providecommand \@@href[1]{\endgroup#1\@@endlink}%
\providecommand \@sanitize@url [0]{\catcode `\\12\catcode `\$12\catcode
  `\&12\catcode `\#12\catcode `\^12\catcode `\_12\catcode `\%12\relax}%
\providecommand \@@startlink[1]{}%
\providecommand \@@endlink[0]{}%
\providecommand \url  [0]{\begingroup\@sanitize@url \@url }%
\providecommand \@url [1]{\endgroup\@href {#1}{\urlprefix }}%
\providecommand \urlprefix  [0]{URL }%
\providecommand \Eprint [0]{\href }%
\providecommand \doibase [0]{http://dx.doi.org/}%
\providecommand \selectlanguage [0]{\@gobble}%
\providecommand \bibinfo  [0]{\@secondoftwo}%
\providecommand \bibfield  [0]{\@secondoftwo}%
\providecommand \translation [1]{[#1]}%
\providecommand \BibitemOpen [0]{}%
\providecommand \bibitemStop [0]{}%
\providecommand \bibitemNoStop [0]{.\EOS\space}%
\providecommand \EOS [0]{\spacefactor3000\relax}%
\providecommand \BibitemShut  [1]{\csname bibitem#1\endcsname}%
\let\auto@bib@innerbib\@empty
%</preamble>
\bibitem [{\citenamefont {Frenkel}\ and\ \citenamefont
  {Smit}(2001)}]{frenkel2001understanding}%
  \BibitemOpen
  \bibfield  {author} {\bibinfo {author} {\bibfnamefont {D.}~\bibnamefont
  {Frenkel}}\ and\ \bibinfo {author} {\bibfnamefont {B.}~\bibnamefont {Smit}},\
  }\href@noop {} {\emph {\bibinfo {title} {Understanding molecular simulation:
  from algorithms to applications}}},\ Vol.~\bibinfo {volume} {1}\ (\bibinfo
  {publisher} {Academic press},\ \bibinfo {year} {2001})\BibitemShut {NoStop}%
\bibitem [{\citenamefont {Torrie}\ and\ \citenamefont
  {Valleau}(1977)}]{torrie1977nonphysical}%
  \BibitemOpen
  \bibfield  {author} {\bibinfo {author} {\bibfnamefont {G.~M.}\ \bibnamefont
  {Torrie}}\ and\ \bibinfo {author} {\bibfnamefont {J.~P.}\ \bibnamefont
  {Valleau}},\ }\href@noop {} {\bibfield  {journal} {\bibinfo  {journal}
  {Journal of Computational Physics}\ }\textbf {\bibinfo {volume} {23}},\
  \bibinfo {pages} {187} (\bibinfo {year} {1977})}\BibitemShut {NoStop}%
\bibitem [{\citenamefont {Allen}\ \emph {et~al.}(2009)\citenamefont {Allen},
  \citenamefont {Valeriani},\ and\ \citenamefont {ten
  Wolde}}]{allen2009forward}%
  \BibitemOpen
  \bibfield  {author} {\bibinfo {author} {\bibfnamefont {R.~J.}\ \bibnamefont
  {Allen}}, \bibinfo {author} {\bibfnamefont {C.}~\bibnamefont {Valeriani}}, \
  and\ \bibinfo {author} {\bibfnamefont {P.~R.}\ \bibnamefont {ten Wolde}},\
  }\href@noop {} {\bibfield  {journal} {\bibinfo  {journal} {Journal of
  physics: Condensed matter}\ }\textbf {\bibinfo {volume} {21}},\ \bibinfo
  {pages} {463102} (\bibinfo {year} {2009})}\BibitemShut {NoStop}%
\bibitem [{\citenamefont {Bolhuis}\ \emph {et~al.}(2002)\citenamefont
  {Bolhuis}, \citenamefont {Chandler}, \citenamefont {Dellago},\ and\
  \citenamefont {Geissler}}]{bolhuis2002transition}%
  \BibitemOpen
  \bibfield  {author} {\bibinfo {author} {\bibfnamefont {P.~G.}\ \bibnamefont
  {Bolhuis}}, \bibinfo {author} {\bibfnamefont {D.}~\bibnamefont {Chandler}},
  \bibinfo {author} {\bibfnamefont {C.}~\bibnamefont {Dellago}}, \ and\
  \bibinfo {author} {\bibfnamefont {P.~L.}\ \bibnamefont {Geissler}},\
  }\href@noop {} {\bibfield  {journal} {\bibinfo  {journal} {Annual Review of
  Physical Chemistry}\ }\textbf {\bibinfo {volume} {53}},\ \bibinfo {pages}
  {291} (\bibinfo {year} {2002})}\BibitemShut {NoStop}%
\bibitem [{\citenamefont {Giardina}\ \emph {et~al.}(2011)\citenamefont
  {Giardina}, \citenamefont {Kurchan}, \citenamefont {Lecomte},\ and\
  \citenamefont {Tailleur}}]{giardina2011simulating}%
  \BibitemOpen
  \bibfield  {author} {\bibinfo {author} {\bibfnamefont {C.}~\bibnamefont
  {Giardina}}, \bibinfo {author} {\bibfnamefont {J.}~\bibnamefont {Kurchan}},
  \bibinfo {author} {\bibfnamefont {V.}~\bibnamefont {Lecomte}}, \ and\
  \bibinfo {author} {\bibfnamefont {J.}~\bibnamefont {Tailleur}},\ }\href@noop
  {} {\bibfield  {journal} {\bibinfo  {journal} {Journal of statistical
  physics}\ }\textbf {\bibinfo {volume} {145}},\ \bibinfo {pages} {787}
  (\bibinfo {year} {2011})}\BibitemShut {NoStop}%
\bibitem [{\citenamefont {Giardina}\ \emph {et~al.}(2006)\citenamefont
  {Giardina}, \citenamefont {Kurchan},\ and\ \citenamefont
  {Peliti}}]{giardina2006direct}%
  \BibitemOpen
  \bibfield  {author} {\bibinfo {author} {\bibfnamefont {C.}~\bibnamefont
  {Giardina}}, \bibinfo {author} {\bibfnamefont {J.}~\bibnamefont {Kurchan}}, \
  and\ \bibinfo {author} {\bibfnamefont {L.}~\bibnamefont {Peliti}},\
  }\href@noop {} {\bibfield  {journal} {\bibinfo  {journal} {Physical Review
  Letters}\ }\textbf {\bibinfo {volume} {96}},\ \bibinfo {pages} {120603}
  (\bibinfo {year} {2006})}\BibitemShut {NoStop}%
\bibitem [{\citenamefont {Nemoto}\ \emph {et~al.}(2016)\citenamefont {Nemoto},
  \citenamefont {Bouchet}, \citenamefont {Jack},\ and\ \citenamefont
  {Lecomte}}]{nemoto2016population}%
  \BibitemOpen
  \bibfield  {author} {\bibinfo {author} {\bibfnamefont {T.}~\bibnamefont
  {Nemoto}}, \bibinfo {author} {\bibfnamefont {F.}~\bibnamefont {Bouchet}},
  \bibinfo {author} {\bibfnamefont {R.~L.}\ \bibnamefont {Jack}}, \ and\
  \bibinfo {author} {\bibfnamefont {V.}~\bibnamefont {Lecomte}},\ }\href@noop
  {} {\bibfield  {journal} {\bibinfo  {journal} {Physical Review E}\ }\textbf
  {\bibinfo {volume} {93}},\ \bibinfo {pages} {062123} (\bibinfo {year}
  {2016})}\BibitemShut {NoStop}%
\bibitem [{\citenamefont {Lecomte}\ and\ \citenamefont
  {Tailleur}(2007)}]{lecomte2007numerical}%
  \BibitemOpen
  \bibfield  {author} {\bibinfo {author} {\bibfnamefont {V.}~\bibnamefont
  {Lecomte}}\ and\ \bibinfo {author} {\bibfnamefont {J.}~\bibnamefont
  {Tailleur}},\ }\href@noop {} {\bibfield  {journal} {\bibinfo  {journal}
  {Journal of Statistical Mechanics: Theory and Experiment}\ }\textbf {\bibinfo
  {volume} {2007}},\ \bibinfo {pages} {P03004} (\bibinfo {year}
  {2007})}\BibitemShut {NoStop}%
\bibitem [{\citenamefont {Nemoto}(2016)}]{nemoto2016iterative}%
  \BibitemOpen
  \bibfield  {author} {\bibinfo {author} {\bibfnamefont {T.}~\bibnamefont
  {Nemoto}},\ }in\ \href@noop {} {\emph {\bibinfo {booktitle} {Phenomenological
  Structure for the Large Deviation Principle in Time-Series Statistics}}}\
  (\bibinfo  {publisher} {Springer},\ \bibinfo {year} {2016})\ pp.\ \bibinfo
  {pages} {17--39}\BibitemShut {NoStop}%
\bibitem [{\citenamefont {Bucklew}(1990)}]{bucklew1990large}%
  \BibitemOpen
  \bibfield  {author} {\bibinfo {author} {\bibfnamefont {J.~A.}\ \bibnamefont
  {Bucklew}},\ }\href@noop {} {\emph {\bibinfo {title} {Large deviation
  techniques in decision, simulation, and estimation}}}\ (\bibinfo  {publisher}
  {Wiley New York},\ \bibinfo {year} {1990})\BibitemShut {NoStop}%
\bibitem [{\citenamefont {Touchette}(2009)}]{touchette2009large}%
  \BibitemOpen
  \bibfield  {author} {\bibinfo {author} {\bibfnamefont {H.}~\bibnamefont
  {Touchette}},\ }\href@noop {} {\bibfield  {journal} {\bibinfo  {journal}
  {Physics Reports}\ }\textbf {\bibinfo {volume} {478}},\ \bibinfo {pages} {1}
  (\bibinfo {year} {2009})}\BibitemShut {NoStop}%
\bibitem [{\citenamefont {Garrahan}\ \emph {et~al.}(2009)\citenamefont
  {Garrahan}, \citenamefont {Jack}, \citenamefont {Lecomte}, \citenamefont
  {Pitard}, \citenamefont {van Duijvendijk},\ and\ \citenamefont {van
  Wijland}}]{garrahan2009first}%
  \BibitemOpen
  \bibfield  {author} {\bibinfo {author} {\bibfnamefont {J.~P.}\ \bibnamefont
  {Garrahan}}, \bibinfo {author} {\bibfnamefont {R.~L.}\ \bibnamefont {Jack}},
  \bibinfo {author} {\bibfnamefont {V.}~\bibnamefont {Lecomte}}, \bibinfo
  {author} {\bibfnamefont {E.}~\bibnamefont {Pitard}}, \bibinfo {author}
  {\bibfnamefont {K.}~\bibnamefont {van Duijvendijk}}, \ and\ \bibinfo {author}
  {\bibfnamefont {F.}~\bibnamefont {van Wijland}},\ }\href@noop {} {\bibfield
  {journal} {\bibinfo  {journal} {Journal of Physics A: Mathematical and
  Theoretical}\ }\textbf {\bibinfo {volume} {42}},\ \bibinfo {pages} {075007}
  (\bibinfo {year} {2009})}\BibitemShut {NoStop}%
\bibitem [{\citenamefont {Maes}\ and\ \citenamefont
  {Neto{\v{c}}n{\`y}}(2008)}]{maes2008canonical}%
  \BibitemOpen
  \bibfield  {author} {\bibinfo {author} {\bibfnamefont {C.}~\bibnamefont
  {Maes}}\ and\ \bibinfo {author} {\bibfnamefont {K.}~\bibnamefont
  {Neto{\v{c}}n{\`y}}},\ }\href@noop {} {\bibfield  {journal} {\bibinfo
  {journal} {EPL (Europhysics Letters)}\ }\textbf {\bibinfo {volume} {82}},\
  \bibinfo {pages} {30003} (\bibinfo {year} {2008})}\BibitemShut {NoStop}%
\bibitem [{\citenamefont {Jack}\ and\ \citenamefont
  {Sollich}(2015)}]{jack2015effective}%
  \BibitemOpen
  \bibfield  {author} {\bibinfo {author} {\bibfnamefont {R.~L.}\ \bibnamefont
  {Jack}}\ and\ \bibinfo {author} {\bibfnamefont {P.}~\bibnamefont {Sollich}},\
  }\href@noop {} {\bibfield  {journal} {\bibinfo  {journal} {The European
  Physical Journal Special Topics}\ }\textbf {\bibinfo {volume} {224}},\
  \bibinfo {pages} {2351} (\bibinfo {year} {2015})}\BibitemShut {NoStop}%
\bibitem [{\citenamefont {Nemoto}\ and\ \citenamefont
  {Sasa}(2014)}]{nemoto2014computation}%
  \BibitemOpen
  \bibfield  {author} {\bibinfo {author} {\bibfnamefont {T.}~\bibnamefont
  {Nemoto}}\ and\ \bibinfo {author} {\bibfnamefont {S.-i.}\ \bibnamefont
  {Sasa}},\ }\href@noop {} {\bibfield  {journal} {\bibinfo  {journal} {Physical
  Review Letters}\ }\textbf {\bibinfo {volume} {112}},\ \bibinfo {pages}
  {090602} (\bibinfo {year} {2014})}\BibitemShut {NoStop}%
\bibitem [{\citenamefont {Klymko}\ \emph {et~al.}(2017)\citenamefont {Klymko},
  \citenamefont {Geissler}, \citenamefont {Garrahan},\ and\ \citenamefont
  {Whitelam}}]{klymko2017rare}%
  \BibitemOpen
  \bibfield  {author} {\bibinfo {author} {\bibfnamefont {K.}~\bibnamefont
  {Klymko}}, \bibinfo {author} {\bibfnamefont {P.~L.}\ \bibnamefont
  {Geissler}}, \bibinfo {author} {\bibfnamefont {J.~P.}\ \bibnamefont
  {Garrahan}}, \ and\ \bibinfo {author} {\bibfnamefont {S.}~\bibnamefont
  {Whitelam}},\ }\href@noop {} {\bibfield  {journal} {\bibinfo  {journal}
  {arXiv preprint arXiv:1707.00767}\ } (\bibinfo {year} {2017})}\BibitemShut
  {NoStop}%
\bibitem [{\citenamefont {Whitelam}\ \emph {et~al.}(2017)\citenamefont
  {Whitelam}, \citenamefont {Klymko},\ and\ \citenamefont {Mandal}}]{lattice1}%
  \BibitemOpen
  \bibfield  {author} {\bibinfo {author} {\bibfnamefont {S.}~\bibnamefont
  {Whitelam}}, \bibinfo {author} {\bibfnamefont {K.}~\bibnamefont {Klymko}}, \
  and\ \bibinfo {author} {\bibfnamefont {D.}~\bibnamefont {Mandal}},\
  }\href@noop {} {\bibfield  {journal} {\bibinfo  {journal} {arXiv preprint
  arXiv:1709.03951}\ } (\bibinfo {year} {2017})}\BibitemShut {NoStop}%
\bibitem [{\citenamefont {Evans}(2004{\natexlab{a}})}]{evans2004rules}%
  \BibitemOpen
  \bibfield  {author} {\bibinfo {author} {\bibfnamefont {R.}~\bibnamefont
  {Evans}},\ }\href@noop {} {\bibfield  {journal} {\bibinfo  {journal}
  {Physical Review Letters}\ }\textbf {\bibinfo {volume} {92}},\ \bibinfo
  {pages} {150601} (\bibinfo {year} {2004}{\natexlab{a}})}\BibitemShut
  {NoStop}%
\bibitem [{\citenamefont {Evans}(2004{\natexlab{b}})}]{evans2004detailed}%
  \BibitemOpen
  \bibfield  {author} {\bibinfo {author} {\bibfnamefont {R.}~\bibnamefont
  {Evans}},\ }\href@noop {} {\bibfield  {journal} {\bibinfo  {journal} {Journal
  of Physics A: Mathematical and General}\ }\textbf {\bibinfo {volume} {38}},\
  \bibinfo {pages} {293} (\bibinfo {year} {2004}{\natexlab{b}})}\BibitemShut
  {NoStop}%
\bibitem [{\citenamefont {Chetrite}\ and\ \citenamefont
  {Touchette}(2013)}]{chetrite2013nonequilibrium}%
  \BibitemOpen
  \bibfield  {author} {\bibinfo {author} {\bibfnamefont {R.}~\bibnamefont
  {Chetrite}}\ and\ \bibinfo {author} {\bibfnamefont {H.}~\bibnamefont
  {Touchette}},\ }\href@noop {} {\bibfield  {journal} {\bibinfo  {journal}
  {Physical Review Letters}\ }\textbf {\bibinfo {volume} {111}},\ \bibinfo
  {pages} {120601} (\bibinfo {year} {2013})}\BibitemShut {NoStop}%
\bibitem [{\citenamefont {Chetrite}\ and\ \citenamefont
  {Touchette}(2014)}]{Chetrite2014}%
  \BibitemOpen
  \bibfield  {author} {\bibinfo {author} {\bibfnamefont {R.}~\bibnamefont
  {Chetrite}}\ and\ \bibinfo {author} {\bibfnamefont {H.}~\bibnamefont
  {Touchette}},\ }\bibfield  {booktitle} {\emph {\bibinfo {booktitle} {Annales
  Henri Poincar{\'e}}},\ }\href@noop {} {\bibfield  {journal} {\bibinfo
  {journal} {Ann. Henri Poincar{\'e}}\ }\textbf {\bibinfo {volume} {16}},\
  \bibinfo {pages} {1} (\bibinfo {year} {2014})}\BibitemShut {NoStop}%
\bibitem [{\citenamefont {Dinwoodie}\ \emph {et~al.}(1992)\citenamefont
  {Dinwoodie}, \citenamefont {Zabell} \emph {et~al.}}]{dinwoodie1992large}%
  \BibitemOpen
  \bibfield  {author} {\bibinfo {author} {\bibfnamefont {I.~H.}\ \bibnamefont
  {Dinwoodie}}, \bibinfo {author} {\bibfnamefont {S.~L.}\ \bibnamefont
  {Zabell}},  \emph {et~al.},\ }\href@noop {} {\bibfield  {journal} {\bibinfo
  {journal} {The Annals of Probability}\ }\textbf {\bibinfo {volume} {20}},\
  \bibinfo {pages} {1147} (\bibinfo {year} {1992})}\BibitemShut {NoStop}%
\bibitem [{\citenamefont {Dinwoodie}(1993)}]{dinwoodie1993identifying}%
  \BibitemOpen
  \bibfield  {author} {\bibinfo {author} {\bibfnamefont {I.~H.}\ \bibnamefont
  {Dinwoodie}},\ }\href@noop {} {\bibfield  {journal} {\bibinfo  {journal}
  {Annals of probability}\ }\textbf {\bibinfo {volume} {21}},\ \bibinfo {pages}
  {216} (\bibinfo {year} {1993})}\BibitemShut {NoStop}%
\bibitem [{\citenamefont {Rohwer}\ \emph {et~al.}(2015)\citenamefont {Rohwer},
  \citenamefont {Angeletti},\ and\ \citenamefont
  {Touchette}}]{rohwer2015convergence}%
  \BibitemOpen
  \bibfield  {author} {\bibinfo {author} {\bibfnamefont {C.~M.}\ \bibnamefont
  {Rohwer}}, \bibinfo {author} {\bibfnamefont {F.}~\bibnamefont {Angeletti}}, \
  and\ \bibinfo {author} {\bibfnamefont {H.}~\bibnamefont {Touchette}},\
  }\href@noop {} {\bibfield  {journal} {\bibinfo  {journal} {Physical Review
  E}\ }\textbf {\bibinfo {volume} {92}},\ \bibinfo {pages} {052104} (\bibinfo
  {year} {2015})}\BibitemShut {NoStop}%
\bibitem [{\citenamefont {Touchette}(2011)}]{touchette2011basic}%
  \BibitemOpen
  \bibfield  {author} {\bibinfo {author} {\bibfnamefont {H.}~\bibnamefont
  {Touchette}},\ }in\ \href@noop {} {\emph {\bibinfo {booktitle} {Modern
  Computational Science 11: Lecture Notes from the 3rd International Oldenburg
  Summer School}}},\ \bibinfo {editor} {edited by\ \bibinfo {editor}
  {\bibfnamefont {R.}~\bibnamefont {Leidl}}\ and\ \bibinfo {editor}
  {\bibfnamefont {A.~K.}\ \bibnamefont {Hartmann}}}\ (\bibinfo  {publisher}
  {BIS-Verlag der Carl von Ossietzky Universit\"at Oldenburg},\ \bibinfo {year}
  {2011})\BibitemShut {NoStop}%
\bibitem [{\citenamefont {Lewis}\ and\ \citenamefont
  {Russell}(1997)}]{lewis1997introduction}%
  \BibitemOpen
  \bibfield  {author} {\bibinfo {author} {\bibfnamefont {J.~T.}\ \bibnamefont
  {Lewis}}\ and\ \bibinfo {author} {\bibfnamefont {R.}~\bibnamefont
  {Russell}},\ }\href@noop {} {\bibfield  {journal} {\bibinfo  {journal}
  {Dublin Institute for Advanced Studies}\ ,\ \bibinfo {pages} {1}} (\bibinfo
  {year} {1997})}\BibitemShut {NoStop}%
\bibitem [{\citenamefont {Gillespie}(2005)}]{gillespie2005general}%
  \BibitemOpen
  \bibfield  {author} {\bibinfo {author} {\bibfnamefont {D.}~\bibnamefont
  {Gillespie}},\ }\href@noop {} {\bibfield  {journal} {\bibinfo  {journal}
  {Journal of Computational Physics}\ }\textbf {\bibinfo {volume} {22}},\
  \bibinfo {pages} {403} (\bibinfo {year} {2005})}\BibitemShut {NoStop}%
\bibitem [{\citenamefont {Oono}(1989)}]{oono1989large}%
  \BibitemOpen
  \bibfield  {author} {\bibinfo {author} {\bibfnamefont {Y.}~\bibnamefont
  {Oono}},\ }\href@noop {} {\bibfield  {journal} {\bibinfo  {journal} {Progress
  of Theoretical Physics Supplement}\ }\textbf {\bibinfo {volume} {99}},\
  \bibinfo {pages} {165} (\bibinfo {year} {1989})}\BibitemShut {NoStop}%
\bibitem [{\citenamefont {Pietzonka}\ \emph {et~al.}(2016)\citenamefont
  {Pietzonka}, \citenamefont {Kleinbeck},\ and\ \citenamefont
  {Seifert}}]{pietzonka2016extreme}%
  \BibitemOpen
  \bibfield  {author} {\bibinfo {author} {\bibfnamefont {P.}~\bibnamefont
  {Pietzonka}}, \bibinfo {author} {\bibfnamefont {K.}~\bibnamefont
  {Kleinbeck}}, \ and\ \bibinfo {author} {\bibfnamefont {U.}~\bibnamefont
  {Seifert}},\ }\href@noop {} {\bibfield  {journal} {\bibinfo  {journal} {New
  Journal of Physics}\ }\textbf {\bibinfo {volume} {18}},\ \bibinfo {pages}
  {052001} (\bibinfo {year} {2016})}\BibitemShut {NoStop}%
\bibitem [{\citenamefont {Thompson}\ \emph {et~al.}(2011)\citenamefont
  {Thompson}, \citenamefont {Tailleur}, \citenamefont {Cates},\ and\
  \citenamefont {Blythe}}]{thompson2011lattice}%
  \BibitemOpen
  \bibfield  {author} {\bibinfo {author} {\bibfnamefont {A.}~\bibnamefont
  {Thompson}}, \bibinfo {author} {\bibfnamefont {J.}~\bibnamefont {Tailleur}},
  \bibinfo {author} {\bibfnamefont {M.}~\bibnamefont {Cates}}, \ and\ \bibinfo
  {author} {\bibfnamefont {R.}~\bibnamefont {Blythe}},\ }\href@noop {}
  {\bibfield  {journal} {\bibinfo  {journal} {Journal of Statistical Mechanics:
  Theory and Experiment}\ }\textbf {\bibinfo {volume} {2011}},\ \bibinfo
  {pages} {P02029} (\bibinfo {year} {2011})}\BibitemShut {NoStop}%
\bibitem [{\citenamefont {Tailleur}\ and\ \citenamefont
  {Cates}(2008)}]{tailleur2008statistical}%
  \BibitemOpen
  \bibfield  {author} {\bibinfo {author} {\bibfnamefont {J.}~\bibnamefont
  {Tailleur}}\ and\ \bibinfo {author} {\bibfnamefont {M.}~\bibnamefont
  {Cates}},\ }\href@noop {} {\bibfield  {journal} {\bibinfo  {journal}
  {Physical Review Letters}\ }\textbf {\bibinfo {volume} {100}},\ \bibinfo
  {pages} {218103} (\bibinfo {year} {2008})}\BibitemShut {NoStop}%
\bibitem [{\citenamefont {Weitz}\ \emph {et~al.}(2015)\citenamefont {Weitz},
  \citenamefont {Deutsch},\ and\ \citenamefont {Peruani}}]{weitz2015self}%
  \BibitemOpen
  \bibfield  {author} {\bibinfo {author} {\bibfnamefont {S.}~\bibnamefont
  {Weitz}}, \bibinfo {author} {\bibfnamefont {A.}~\bibnamefont {Deutsch}}, \
  and\ \bibinfo {author} {\bibfnamefont {F.}~\bibnamefont {Peruani}},\
  }\href@noop {} {\bibfield  {journal} {\bibinfo  {journal} {Physical Review
  E}\ }\textbf {\bibinfo {volume} {92}},\ \bibinfo {pages} {012322} (\bibinfo
  {year} {2015})}\BibitemShut {NoStop}%
\bibitem [{\citenamefont {Baskaran}\ and\ \citenamefont
  {Marchetti}(2009)}]{baskaran2009statistical}%
  \BibitemOpen
  \bibfield  {author} {\bibinfo {author} {\bibfnamefont {A.}~\bibnamefont
  {Baskaran}}\ and\ \bibinfo {author} {\bibfnamefont {M.~C.}\ \bibnamefont
  {Marchetti}},\ }\href@noop {} {\bibfield  {journal} {\bibinfo  {journal}
  {Proceedings of the National Academy of Sciences}\ }\textbf {\bibinfo
  {volume} {106}},\ \bibinfo {pages} {15567} (\bibinfo {year}
  {2009})}\BibitemShut {NoStop}%
\bibitem [{\citenamefont {Peruani}\ \emph {et~al.}(2012)\citenamefont
  {Peruani}, \citenamefont {Starru{\ss}}, \citenamefont {Jakovljevic},
  \citenamefont {S{\o}gaard-Andersen}, \citenamefont {Deutsch},\ and\
  \citenamefont {B{\"a}r}}]{peruani2012collective}%
  \BibitemOpen
  \bibfield  {author} {\bibinfo {author} {\bibfnamefont {F.}~\bibnamefont
  {Peruani}}, \bibinfo {author} {\bibfnamefont {J.}~\bibnamefont
  {Starru{\ss}}}, \bibinfo {author} {\bibfnamefont {V.}~\bibnamefont
  {Jakovljevic}}, \bibinfo {author} {\bibfnamefont {L.}~\bibnamefont
  {S{\o}gaard-Andersen}}, \bibinfo {author} {\bibfnamefont {A.}~\bibnamefont
  {Deutsch}}, \ and\ \bibinfo {author} {\bibfnamefont {M.}~\bibnamefont
  {B{\"a}r}},\ }\href@noop {} {\bibfield  {journal} {\bibinfo  {journal}
  {Physical Review Letters}\ }\textbf {\bibinfo {volume} {108}},\ \bibinfo
  {pages} {098102} (\bibinfo {year} {2012})}\BibitemShut {NoStop}%
\bibitem [{\citenamefont {Barr{\'e}}\ \emph {et~al.}(2015)\citenamefont
  {Barr{\'e}}, \citenamefont {Ch{\'e}trite}, \citenamefont {Muratori},\ and\
  \citenamefont {Peruani}}]{barre2015motility}%
  \BibitemOpen
  \bibfield  {author} {\bibinfo {author} {\bibfnamefont {J.}~\bibnamefont
  {Barr{\'e}}}, \bibinfo {author} {\bibfnamefont {R.}~\bibnamefont
  {Ch{\'e}trite}}, \bibinfo {author} {\bibfnamefont {M.}~\bibnamefont
  {Muratori}}, \ and\ \bibinfo {author} {\bibfnamefont {F.}~\bibnamefont
  {Peruani}},\ }\href@noop {} {\bibfield  {journal} {\bibinfo  {journal}
  {Journal of Statistical Physics}\ }\textbf {\bibinfo {volume} {158}},\
  \bibinfo {pages} {589} (\bibinfo {year} {2015})}\BibitemShut {NoStop}%
\bibitem [{\citenamefont {Peruani}\ \emph {et~al.}(2011)\citenamefont
  {Peruani}, \citenamefont {Klauss}, \citenamefont {Deutsch},\ and\
  \citenamefont {Voss-Boehme}}]{peruani2011traffic}%
  \BibitemOpen
  \bibfield  {author} {\bibinfo {author} {\bibfnamefont {F.}~\bibnamefont
  {Peruani}}, \bibinfo {author} {\bibfnamefont {T.}~\bibnamefont {Klauss}},
  \bibinfo {author} {\bibfnamefont {A.}~\bibnamefont {Deutsch}}, \ and\
  \bibinfo {author} {\bibfnamefont {A.}~\bibnamefont {Voss-Boehme}},\
  }\href@noop {} {\bibfield  {journal} {\bibinfo  {journal} {Physical Review
  Letters}\ }\textbf {\bibinfo {volume} {106}},\ \bibinfo {pages} {128101}
  (\bibinfo {year} {2011})}\BibitemShut {NoStop}%
\bibitem [{\citenamefont {Ramaswamy}(2010)}]{ramaswamy2010mechanics}%
  \BibitemOpen
  \bibfield  {author} {\bibinfo {author} {\bibfnamefont {S.}~\bibnamefont
  {Ramaswamy}},\ }\href@noop {} {\bibfield  {journal} {\bibinfo  {journal} {The
  Annual Review of Condensed Matter Physics is}\ }\textbf {\bibinfo {volume}
  {1}},\ \bibinfo {pages} {323} (\bibinfo {year} {2010})}\BibitemShut {NoStop}%
\bibitem [{\citenamefont {Romanczuk}\ \emph {et~al.}(2012)\citenamefont
  {Romanczuk}, \citenamefont {Bar}, \citenamefont {Ebeling}, \citenamefont
  {Lindner},\ and\ \citenamefont {Schimansky-Geier}}]{Romanczuk_2012}%
  \BibitemOpen
  \bibfield  {author} {\bibinfo {author} {\bibfnamefont {P.}~\bibnamefont
  {Romanczuk}}, \bibinfo {author} {\bibfnamefont {M.}~\bibnamefont {Bar}},
  \bibinfo {author} {\bibfnamefont {W.}~\bibnamefont {Ebeling}}, \bibinfo
  {author} {\bibfnamefont {B.}~\bibnamefont {Lindner}}, \ and\ \bibinfo
  {author} {\bibfnamefont {L.}~\bibnamefont {Schimansky-Geier}},\ }\href@noop
  {} {\bibfield  {journal} {\bibinfo  {journal} {European Physical Journal
  Special Topics}\ }\textbf {\bibinfo {volume} {202}},\ \bibinfo {pages} {1}
  (\bibinfo {year} {2012})}\BibitemShut {NoStop}%
\bibitem [{\citenamefont {Redner}\ \emph {et~al.}(2013)\citenamefont {Redner},
  \citenamefont {Hagan},\ and\ \citenamefont {Baskaran}}]{redner2013structure}%
  \BibitemOpen
  \bibfield  {author} {\bibinfo {author} {\bibfnamefont {G.~S.}\ \bibnamefont
  {Redner}}, \bibinfo {author} {\bibfnamefont {M.~F.}\ \bibnamefont {Hagan}}, \
  and\ \bibinfo {author} {\bibfnamefont {A.}~\bibnamefont {Baskaran}},\
  }\href@noop {} {\bibfield  {journal} {\bibinfo  {journal} {Physical Review
  Letters}\ }\textbf {\bibinfo {volume} {110}},\ \bibinfo {pages} {055701}
  (\bibinfo {year} {2013})}\BibitemShut {NoStop}%
\bibitem [{\citenamefont {Marchetti}\ \emph {et~al.}(2013)\citenamefont
  {Marchetti}, \citenamefont {Joanny}, \citenamefont {Ramaswamy}, \citenamefont
  {Liverpool}, \citenamefont {Prost}, \citenamefont {Rao},\ and\ \citenamefont
  {Simha}}]{marchetti2013hydrodynamics}%
  \BibitemOpen
  \bibfield  {author} {\bibinfo {author} {\bibfnamefont {M.~C.}\ \bibnamefont
  {Marchetti}}, \bibinfo {author} {\bibfnamefont {J.}~\bibnamefont {Joanny}},
  \bibinfo {author} {\bibfnamefont {S.}~\bibnamefont {Ramaswamy}}, \bibinfo
  {author} {\bibfnamefont {T.}~\bibnamefont {Liverpool}}, \bibinfo {author}
  {\bibfnamefont {J.}~\bibnamefont {Prost}}, \bibinfo {author} {\bibfnamefont
  {M.}~\bibnamefont {Rao}}, \ and\ \bibinfo {author} {\bibfnamefont {R.~A.}\
  \bibnamefont {Simha}},\ }\href@noop {} {\bibfield  {journal} {\bibinfo
  {journal} {Reviews of Modern Physics}\ }\textbf {\bibinfo {volume} {85}},\
  \bibinfo {pages} {1143} (\bibinfo {year} {2013})}\BibitemShut {NoStop}%
\bibitem [{\citenamefont {Buttinoni}\ \emph {et~al.}(2013)\citenamefont
  {Buttinoni}, \citenamefont {Bialk{\'e}}, \citenamefont {K{\"u}mmel},
  \citenamefont {L{\"o}wen}, \citenamefont {Bechinger},\ and\ \citenamefont
  {Speck}}]{buttinoni2013dynamical}%
  \BibitemOpen
  \bibfield  {author} {\bibinfo {author} {\bibfnamefont {I.}~\bibnamefont
  {Buttinoni}}, \bibinfo {author} {\bibfnamefont {J.}~\bibnamefont
  {Bialk{\'e}}}, \bibinfo {author} {\bibfnamefont {F.}~\bibnamefont
  {K{\"u}mmel}}, \bibinfo {author} {\bibfnamefont {H.}~\bibnamefont
  {L{\"o}wen}}, \bibinfo {author} {\bibfnamefont {C.}~\bibnamefont
  {Bechinger}}, \ and\ \bibinfo {author} {\bibfnamefont {T.}~\bibnamefont
  {Speck}},\ }\href@noop {} {\bibfield  {journal} {\bibinfo  {journal}
  {Physical Review Letters}\ }\textbf {\bibinfo {volume} {110}},\ \bibinfo
  {pages} {238301} (\bibinfo {year} {2013})}\BibitemShut {NoStop}%
\bibitem [{\citenamefont {Stenhammar}\ \emph {et~al.}(2013)\citenamefont
  {Stenhammar}, \citenamefont {Tiribocchi}, \citenamefont {Allen},
  \citenamefont {Marenduzzo},\ and\ \citenamefont
  {Cates}}]{stenhammar2013continuum}%
  \BibitemOpen
  \bibfield  {author} {\bibinfo {author} {\bibfnamefont {J.}~\bibnamefont
  {Stenhammar}}, \bibinfo {author} {\bibfnamefont {A.}~\bibnamefont
  {Tiribocchi}}, \bibinfo {author} {\bibfnamefont {R.~J.}\ \bibnamefont
  {Allen}}, \bibinfo {author} {\bibfnamefont {D.}~\bibnamefont {Marenduzzo}}, \
  and\ \bibinfo {author} {\bibfnamefont {M.~E.}\ \bibnamefont {Cates}},\
  }\href@noop {} {\bibfield  {journal} {\bibinfo  {journal} {Physical Review
  Letters}\ }\textbf {\bibinfo {volume} {111}},\ \bibinfo {pages} {145702}
  (\bibinfo {year} {2013})}\BibitemShut {NoStop}%
\bibitem [{\citenamefont {Yeomans}\ \emph {et~al.}(2014)\citenamefont
  {Yeomans}, \citenamefont {Pushkin},\ and\ \citenamefont
  {Shum}}]{Yeomans_2014}%
  \BibitemOpen
  \bibfield  {author} {\bibinfo {author} {\bibfnamefont {J.}~\bibnamefont
  {Yeomans}}, \bibinfo {author} {\bibfnamefont {D.}~\bibnamefont {Pushkin}}, \
  and\ \bibinfo {author} {\bibfnamefont {H.}~\bibnamefont {Shum}},\ }\href@noop
  {} {\bibfield  {journal} {\bibinfo  {journal} {European Physical Journal
  Special Topics}\ }\textbf {\bibinfo {volume} {223}},\ \bibinfo {pages} {1771}
  (\bibinfo {year} {2014})}\BibitemShut {NoStop}%
\bibitem [{\citenamefont {Cates}\ and\ \citenamefont
  {Tailleur}(2015)}]{cates2015motility}%
  \BibitemOpen
  \bibfield  {author} {\bibinfo {author} {\bibfnamefont {M.~E.}\ \bibnamefont
  {Cates}}\ and\ \bibinfo {author} {\bibfnamefont {J.}~\bibnamefont
  {Tailleur}},\ }\href@noop {} {\bibfield  {journal} {\bibinfo  {journal}
  {Annu. Rev. Condens. Matter Phys.}\ }\textbf {\bibinfo {volume} {6}},\
  \bibinfo {pages} {219} (\bibinfo {year} {2015})}\BibitemShut {NoStop}%
\bibitem [{\citenamefont {Menzel}(2015)}]{Menzel_2015}%
  \BibitemOpen
  \bibfield  {author} {\bibinfo {author} {\bibfnamefont {A.~M.}\ \bibnamefont
  {Menzel}},\ }\href@noop {} {\bibfield  {journal} {\bibinfo  {journal}
  {Physics Reports}\ }\textbf {\bibinfo {volume} {554}},\ \bibinfo {pages} {1}
  (\bibinfo {year} {2015})}\BibitemShut {NoStop}%
\bibitem [{\citenamefont {Bechinger}\ \emph {et~al.}(2016)\citenamefont
  {Bechinger}, \citenamefont {Leonardo}, \citenamefont {Lowen}, \citenamefont
  {Reichhardt}, \citenamefont {Volpe},\ and\ \citenamefont
  {Volpe}}]{Bechinger_2016}%
  \BibitemOpen
  \bibfield  {author} {\bibinfo {author} {\bibfnamefont {C.}~\bibnamefont
  {Bechinger}}, \bibinfo {author} {\bibfnamefont {R.~D.}\ \bibnamefont
  {Leonardo}}, \bibinfo {author} {\bibfnamefont {H.}~\bibnamefont {Lowen}},
  \bibinfo {author} {\bibfnamefont {C.}~\bibnamefont {Reichhardt}}, \bibinfo
  {author} {\bibfnamefont {G.}~\bibnamefont {Volpe}}, \ and\ \bibinfo {author}
  {\bibfnamefont {G.}~\bibnamefont {Volpe}},\ }\href@noop {} {\bibfield
  {journal} {\bibinfo  {journal} {Reviews of Modern Physics}\ }\textbf
  {\bibinfo {volume} {88}},\ \bibinfo {pages} {045006} (\bibinfo {year}
  {2016})}\BibitemShut {NoStop}%
\bibitem [{\citenamefont {Fily}\ and\ \citenamefont
  {Marchetti}(2012)}]{Fily_2012}%
  \BibitemOpen
  \bibfield  {author} {\bibinfo {author} {\bibfnamefont {Y.}~\bibnamefont
  {Fily}}\ and\ \bibinfo {author} {\bibfnamefont {M.~C.}\ \bibnamefont
  {Marchetti}},\ }\href {http://dx.doi.org/10.1103/PhysRevLett.108.235702}
  {\bibfield  {journal} {\bibinfo  {journal} {Physical Review Letters}\
  }\textbf {\bibinfo {volume} {108}},\ \bibinfo {pages} {235702} (\bibinfo
  {year} {2012})}\BibitemShut {NoStop}%
\bibitem [{\citenamefont {Speck}\ \emph {et~al.}(2014)\citenamefont {Speck},
  \citenamefont {Bialk{\'e}}, \citenamefont {Menzel},\ and\ \citenamefont
  {L{\"o}wen}}]{Speck_2014}%
  \BibitemOpen
  \bibfield  {author} {\bibinfo {author} {\bibfnamefont {T.}~\bibnamefont
  {Speck}}, \bibinfo {author} {\bibfnamefont {J.}~\bibnamefont {Bialk{\'e}}},
  \bibinfo {author} {\bibfnamefont {A.~M.}\ \bibnamefont {Menzel}}, \ and\
  \bibinfo {author} {\bibfnamefont {H.}~\bibnamefont {L{\"o}wen}},\ }\href
  {http://dx.doi.org/10.1103/PhysRevLett.112.218304} {\bibfield  {journal}
  {\bibinfo  {journal} {Physical Review Letters}\ }\textbf {\bibinfo {volume}
  {112}},\ \bibinfo {pages} {218304} (\bibinfo {year} {2014})}\BibitemShut
  {NoStop}%
\bibitem [{\citenamefont {Solon}\ and\ \citenamefont
  {Tailleur}(2015)}]{solon2015flocking}%
  \BibitemOpen
  \bibfield  {author} {\bibinfo {author} {\bibfnamefont {A.~P.}\ \bibnamefont
  {Solon}}\ and\ \bibinfo {author} {\bibfnamefont {J.}~\bibnamefont
  {Tailleur}},\ }\href@noop {} {\bibfield  {journal} {\bibinfo  {journal}
  {Physical Review E}\ }\textbf {\bibinfo {volume} {92}},\ \bibinfo {pages}
  {042119} (\bibinfo {year} {2015})}\BibitemShut {NoStop}%
\bibitem [{\citenamefont {Solon}\ and\ \citenamefont
  {Tailleur}(2013)}]{solon2013revisiting}%
  \BibitemOpen
  \bibfield  {author} {\bibinfo {author} {\bibfnamefont {A.}~\bibnamefont
  {Solon}}\ and\ \bibinfo {author} {\bibfnamefont {J.}~\bibnamefont
  {Tailleur}},\ }\href@noop {} {\bibfield  {journal} {\bibinfo  {journal}
  {Physical Review Letters}\ }\textbf {\bibinfo {volume} {111}},\ \bibinfo
  {pages} {078101} (\bibinfo {year} {2013})}\BibitemShut {NoStop}%
\bibitem [{\citenamefont {Pilkiewicz}\ and\ \citenamefont
  {Eaves}(2014)}]{PhysRevE.89.012718}%
  \BibitemOpen
  \bibfield  {author} {\bibinfo {author} {\bibfnamefont {K.~R.}\ \bibnamefont
  {Pilkiewicz}}\ and\ \bibinfo {author} {\bibfnamefont {J.~D.}\ \bibnamefont
  {Eaves}},\ }\href {\doibase 10.1103/PhysRevE.89.012718} {\bibfield  {journal}
  {\bibinfo  {journal} {Phys. Rev. E}\ }\textbf {\bibinfo {volume} {89}},\
  \bibinfo {pages} {012718} (\bibinfo {year} {2014})}\BibitemShut {NoStop}%
\bibitem [{\citenamefont {Peruani}\ \emph {et~al.}(2010)\citenamefont
  {Peruani}, \citenamefont {Schimansky-Geier},\ and\ \citenamefont
  {B{\"a}r}}]{peruani2010cluster}%
  \BibitemOpen
  \bibfield  {author} {\bibinfo {author} {\bibfnamefont {F.}~\bibnamefont
  {Peruani}}, \bibinfo {author} {\bibfnamefont {L.}~\bibnamefont
  {Schimansky-Geier}}, \ and\ \bibinfo {author} {\bibfnamefont
  {M.}~\bibnamefont {B{\"a}r}},\ }\href@noop {} {\bibfield  {journal} {\bibinfo
   {journal} {The European Physical Journal-Special Topics}\ }\textbf {\bibinfo
  {volume} {191}},\ \bibinfo {pages} {173} (\bibinfo {year}
  {2010})}\BibitemShut {NoStop}%
\bibitem [{\citenamefont {Hedges}\ \emph {et~al.}(2009)\citenamefont {Hedges},
  \citenamefont {Jack}, \citenamefont {Garrahan},\ and\ \citenamefont
  {Chandler}}]{hedges2009dynamic}%
  \BibitemOpen
  \bibfield  {author} {\bibinfo {author} {\bibfnamefont {L.~O.}\ \bibnamefont
  {Hedges}}, \bibinfo {author} {\bibfnamefont {R.~L.}\ \bibnamefont {Jack}},
  \bibinfo {author} {\bibfnamefont {J.~P.}\ \bibnamefont {Garrahan}}, \ and\
  \bibinfo {author} {\bibfnamefont {D.}~\bibnamefont {Chandler}},\ }\href@noop
  {} {\bibfield  {journal} {\bibinfo  {journal} {Science}\ }\textbf {\bibinfo
  {volume} {323}},\ \bibinfo {pages} {1309} (\bibinfo {year}
  {2009})}\BibitemShut {NoStop}%
\bibitem [{\citenamefont {Budini}\ \emph {et~al.}(2014)\citenamefont {Budini},
  \citenamefont {Turner},\ and\ \citenamefont
  {Garrahan}}]{budini2014fluctuating}%
  \BibitemOpen
  \bibfield  {author} {\bibinfo {author} {\bibfnamefont {A.~A.}\ \bibnamefont
  {Budini}}, \bibinfo {author} {\bibfnamefont {R.~M.}\ \bibnamefont {Turner}},
  \ and\ \bibinfo {author} {\bibfnamefont {J.~P.}\ \bibnamefont {Garrahan}},\
  }\href@noop {} {\bibfield  {journal} {\bibinfo  {journal} {Journal of
  Statistical Mechanics: Theory and Experiment}\ }\textbf {\bibinfo {volume}
  {2014}},\ \bibinfo {pages} {P03012} (\bibinfo {year} {2014})}\BibitemShut
  {NoStop}%
\bibitem [{\citenamefont {Mey}\ \emph {et~al.}(2014)\citenamefont {Mey},
  \citenamefont {Geissler},\ and\ \citenamefont {Garrahan}}]{mey2014rare}%
  \BibitemOpen
  \bibfield  {author} {\bibinfo {author} {\bibfnamefont {A.~S.}\ \bibnamefont
  {Mey}}, \bibinfo {author} {\bibfnamefont {P.~L.}\ \bibnamefont {Geissler}}, \
  and\ \bibinfo {author} {\bibfnamefont {J.~P.}\ \bibnamefont {Garrahan}},\
  }\href@noop {} {\bibfield  {journal} {\bibinfo  {journal} {Physical Review
  E}\ }\textbf {\bibinfo {volume} {89}},\ \bibinfo {pages} {032109} (\bibinfo
  {year} {2014})}\BibitemShut {NoStop}%
\bibitem [{\citenamefont {Ruelle}(2004)}]{ruelle2004thermodynamic}%
  \BibitemOpen
  \bibfield  {author} {\bibinfo {author} {\bibfnamefont {D.}~\bibnamefont
  {Ruelle}},\ }\href@noop {} {\emph {\bibinfo {title} {Thermodynamic formalism:
  the mathematical structure of equilibrium statistical mechanics}}}\ (\bibinfo
   {publisher} {Cambridge University Press},\ \bibinfo {year}
  {2004})\BibitemShut {NoStop}%
\bibitem [{\citenamefont {Garrahan}\ \emph {et~al.}(2007)\citenamefont
  {Garrahan}, \citenamefont {Jack}, \citenamefont {Lecomte}, \citenamefont
  {Pitard}, \citenamefont {van Duijvendijk},\ and\ \citenamefont {van
  Wijland}}]{garrahan2007dynamical}%
  \BibitemOpen
  \bibfield  {author} {\bibinfo {author} {\bibfnamefont {J.~P.}\ \bibnamefont
  {Garrahan}}, \bibinfo {author} {\bibfnamefont {R.~L.}\ \bibnamefont {Jack}},
  \bibinfo {author} {\bibfnamefont {V.}~\bibnamefont {Lecomte}}, \bibinfo
  {author} {\bibfnamefont {E.}~\bibnamefont {Pitard}}, \bibinfo {author}
  {\bibfnamefont {K.}~\bibnamefont {van Duijvendijk}}, \ and\ \bibinfo {author}
  {\bibfnamefont {F.}~\bibnamefont {van Wijland}},\ }\href@noop {} {\bibfield
  {journal} {\bibinfo  {journal} {Physical Review Letters}\ }\textbf {\bibinfo
  {volume} {98}},\ \bibinfo {pages} {195702} (\bibinfo {year}
  {2007})}\BibitemShut {NoStop}%
\bibitem [{\citenamefont {Lecomte}\ \emph {et~al.}(2007)\citenamefont
  {Lecomte}, \citenamefont {Appert-Rolland},\ and\ \citenamefont {van
  Wijland}}]{Lecomte2007}%
  \BibitemOpen
  \bibfield  {author} {\bibinfo {author} {\bibfnamefont {V.}~\bibnamefont
  {Lecomte}}, \bibinfo {author} {\bibfnamefont {C.}~\bibnamefont
  {Appert-Rolland}}, \ and\ \bibinfo {author} {\bibfnamefont {F.}~\bibnamefont
  {van Wijland}},\ }\href@noop {} {\bibfield  {journal} {\bibinfo  {journal}
  {J. Stat. Phys.}\ }\textbf {\bibinfo {volume} {127}},\ \bibinfo {pages} {51}
  (\bibinfo {year} {2007})}\BibitemShut {NoStop}%
\bibitem [{\citenamefont {Chetrite}\ and\ \citenamefont
  {Touchette}(2015)}]{chetrite2015variational}%
  \BibitemOpen
  \bibfield  {author} {\bibinfo {author} {\bibfnamefont {R.}~\bibnamefont
  {Chetrite}}\ and\ \bibinfo {author} {\bibfnamefont {H.}~\bibnamefont
  {Touchette}},\ }\href@noop {} {\bibfield  {journal} {\bibinfo  {journal}
  {Journal of Statistical Mechanics: Theory and Experiment}\ }\textbf {\bibinfo
  {volume} {2015}},\ \bibinfo {pages} {P12001} (\bibinfo {year}
  {2015})}\BibitemShut {NoStop}%
\bibitem [{\citenamefont {Jasiulewicz}\ and\ \citenamefont
  {Kordecki}(2003)}]{jasiulewicz2003convolutions}%
  \BibitemOpen
  \bibfield  {author} {\bibinfo {author} {\bibfnamefont {H.}~\bibnamefont
  {Jasiulewicz}}\ and\ \bibinfo {author} {\bibfnamefont {W.}~\bibnamefont
  {Kordecki}},\ }\href@noop {} {\bibfield  {journal} {\bibinfo  {journal}
  {Demonstratio Mathematica. Warsaw Technical University Institute of
  Mathematics}\ }\textbf {\bibinfo {volume} {36}},\ \bibinfo {pages} {231}
  (\bibinfo {year} {2003})}\BibitemShut {NoStop}%
\bibitem [{\citenamefont {Buchholz}\ \emph {et~al.}(2014)\citenamefont
  {Buchholz}, \citenamefont {Kriege},\ and\ \citenamefont
  {Felko}}]{buchholz2014input}%
  \BibitemOpen
  \bibfield  {author} {\bibinfo {author} {\bibfnamefont {P.}~\bibnamefont
  {Buchholz}}, \bibinfo {author} {\bibfnamefont {J.}~\bibnamefont {Kriege}}, \
  and\ \bibinfo {author} {\bibfnamefont {I.}~\bibnamefont {Felko}},\
  }\href@noop {} {\emph {\bibinfo {title} {Input modeling with phase-type
  distributions and Markov models: theory and applications}}}\ (\bibinfo
  {publisher} {Springer},\ \bibinfo {year} {2014})\BibitemShut {NoStop}%
\end{thebibliography}


 \end{document}